\documentclass[12pt,preprint]{aastex} %one-column, single-spaced document

\newcommand{\kms}{$\rm km\;s^{-1}$}   
   
\newcommand{\MgI}{Mg~{\small I}}   
\newcommand{\OIII}{[O~{\small III}]}   
\newcommand{\aFe}{[$\alpha$/Fe]} 
  
\newcommand{\Hb}{H$\beta$}  
\newcommand{\Fe}{$\langle$Fe$\rangle$} 
\newcommand{\Mgb}{Mg$b$} 
\newcommand{\Mgd}{Mg$_2$}  
 
\slugcomment{To be published in ApJS} 
 
\shorttitle{Spectroscopy of Coma early-type galaxies. IV} 
\shortauthors{Corsini et al.} 
 
\begin{document} 
 
\title{Spatially resolved spectroscopy of Coma cluster early-type 
  galaxies IV. Completing the dataset$^1$} 
\footnotetext[1]{Based on data collected with the Hobby-Eberly Telescope  
  and the 2.4-m Hiltner Telescope} 
 
\author{E. M. Corsini} 
\affil{Dipartimento di Astronomia, Universit\`a di Padova, vicolo 
    dell'Osservatorio 3, 35122 Padova, Italy} 
\email{enricomaria.corsini@unipd.it} 
 
\author{G. Wegner\altaffilmark{2}} 
\affil{Department of Physics and Astronomy, 6127 Wilder Laboratory,  
    Dartmouth College, Hanover, NH 03755-3528, USA} 
\email{gaw@bellz.dartmouth.edu} 
 
\author{R. P. Saglia} 
\affil{Max-Planck-Institut f\"ur extraterrestrische Physik, 
    Giessenbachstra{\ss}e, D-85748 Garching, Germany 
    D-81679 M\"unchen, Germany} 
\email{saglia@mpe.mpg.de} 
 
\author{J. Thomas and R. Bender} 
\affil{Max-Planck-Institut f\"ur extraterrestrische Physik, 
    Giessenbachstra{\ss}e, D-85748 Garching, Germany 
    D-81679 M\"unchen, Germany} 
\affil{Universit\"ats-Sternwarte M\"unchen, Scheinerstra{\ss}e~1, 
    D-81679 M\"unchen, Germany} 
\email{jthomas@mpe.mpg.de, bender@mpe.mpg.de} 
 
\and 
 
\author{D. Thomas} 
\affil{Institute of Cosmology and Gravitation, Mercantile House, Hampshire  
Terrace, University of Portsmouth, Portsmouth, PO1 2EG, UK} 
\email{daniel.thomas@port.ac.uk} 
 
\altaffiltext{2}{Visiting Astronomer, MDM Observatory, Kitt Peak,
  Arizona, operated by a consortium of Dartmouth College, the University
  of Michigan, Columbia University, the Ohio State University, and Ohio
  University.}
 
\begin{abstract}  
The long-slit spectra obtained along the minor axis, offset major axis 
and diagonal axis are presented for 12 E and S0 galaxies of the Coma 
cluster drawn from a magnitude-limited sample studied before. The 
rotation curves, velocity dispersion profiles and the $H_3$ and $H_4$ 
coefficients of the Hermite decomposition of the line of sight 
velocity distribution are derived. The radial profiles of the \Hb , 
Mg, and Fe line strength indices are measured too. 
In addition, the surface photometry of the central regions of a subsample 
of 4 galaxies recently obtained with Hubble Space Telescope is presented. 
The data will be used to construct dynamical models of the galaxies 
and study their stellar populations.  
\end{abstract} 
 

\section{Introduction} 
 
This is the fourth of a series of papers aimed at investigating the
stellar populations and the kinematics of early-type galaxies in the
Coma Cluster. Spanning about 4 dex in the observed radial
variation of the surface density of cluster members
\citep[e.g.][]{kent1982}, the Coma Cluster is the ideal place to
investigate these galaxy properties as a function of the environmental
density in order to test the theories for galaxy formation and
evolution.
 
The sample of 35 E and S0 galaxies of the Coma Cluster is presented in 
the first paper of the series \citep[hereafter Paper I]{mehlert2000} 
along with the photometry and long-slit spectroscopy along their major 
axis. From these spectra the rotation curves, velocity dispersion 
profiles and the $H_3$ and $H_4$ coefficients of the Hermite 
decomposition of the line-of-sight velocity distribution (LOSVD) were 
measured out to 1--3 effective radii with high signal-to-noise ratio 
($S/N$). Moreover, the radial profiles of the \Hb\, Mg, and Fe line 
strength indices were measured too. 
Subsequently, the spectroscopic database was complemented with the 
long-slit spectra obtained along the minor axis, an offset axis 
parallel to the major one and one diagonal axis for 10 objects 
\citep[hereafter Paper II]{wegner2002}. 
 
The central values and major-axis logarithmic gradients for the line
strength indices were derived by \citet[][hereafter Paper
III]{mehlert2003}. This allowed the estimation of the average ages,
metallicities and \aFe\ ratios in the center and at the effective
radius by using stellar population models with variable element
abundance ratios from \citet{dthomas2003}. There is a dichotomy among
the population of S0 galaxies. Some of them are dominated by old
stellar populations and are indistinguishable from E galaxies. The
remaining ones host very young stellar populations; hence they must
have experienced relatively recent star formation episodes. Most
massive galaxies had the shortest star formation timescales and were
the first to form. The absence of age gradients implies that the
stellar populations at different radii formed at a common epoch. The
\aFe\ enhancement is not restricted to galaxy centers but it is a
global phenomenon. Finally, negative metallicity gradients were
measured to be significantly flatter than what is expected from
gaseous monolithic collapse models. This suggests the importance of
mergers in the galaxy formation history.
 
Here the spectroscopic database of Paper I and II is completed with 
the long-slit spectra obtained along the minor axis, offset major axis 
and one diagonal axis for another 12 objects. As done in Paper II, 
these galaxies were selected from the sample of Paper I as the objects 
with the most extended and precise major-axis kinematics and therefore 
best-suited for dynamical modelling, balancing between the number of E 
and S0 galaxies. Moreover, the surface photometry of the central 
regions of a subsample of 4 galaxies recently obtained with Hubble 
Space Telescope (HST) is presented. 

The data shown here and in Paper I will allow the study of the stellar
population gradients for a large number of early-type galaxies
\citep{dthomas2008} in order to investigate possible systematic 
differences between the disk and bulge components of S0 galaxies.
The photometric and kinematic data of the combined dataset allowed the
construction of dynamical models of the objects to study the
properties of the dark matter halos of flattened and rotating E and S0
galaxies \citep{jthomas2005,jthomas2007}. In fact, the implementation
of Schwarzschild's orbit superposition technique for axisymmetric
potentials by \citet{jthomas2004} was used to derive the stellar
mass-to-light ratios and dark matter halo parameters for a subsample
of 17 galaxies. About 10--50 percent of the mass inside the effective
radius is dark with a central density which is at least one order of
magnitude lower than the luminous mass density. The orbital system of
the stars is reasonably close to isotropy, but the distribution
function shows a lot of fine structure. This study was complementary
to the one presented by \citet{gerhard2001} focusing on round and
non-rotating ellipticals.
 
The HST photometry is described in Sect. \ref{sec:photometry}.  The 
spectroscopic galaxy sample, relative observations and data reduction 
are described in Sect. \ref{sec:spectroscopy}. The measured stellar 
kinematics, and line indices are given in Sect. \ref{sec:results}. 
Conclusions are drawn in Sect. \ref{sec:conclusions}.

\section{HST photometry} 
\label{sec:photometry}

As part of the HST Proposal 10884 (P.I. G. Wegner), the galaxies 
GMP~0756, GMP~1176, GMP~1990, and GMP2440 were observed with Wide 
Field Planetary Camera 2 (WFPC2) on board the HST on 18--24 April 
2007. For each galaxy two 300-sec exposures were taken with the filter 
F622W. All exposures were performed with the telescope guiding
in fine lock, which typically gave an rms tracking error of $0.003$
arcsec. The centers of the galaxies were positioned on the Planetary
Camera chip (PC) in order to get the best possible spatial
resolution. This consists of $800\times800$ pixels of
$0.0455\times0.0455$ arcsec$^2$ each, yielding a field of view of
about $36\times36$ arcsec$^2$.
 
In the following we limit our photometric analysis to the PC chip, 
since we match the nuclear surface brightness profiles to available 
radially extended ground-based photometry. 
 
The images were reduced using the standard reduction pipeline 
maintained by the Space Telescope Science Institute. Reduction steps 
include bias subtraction, dark current subtraction, and flat-fielding 
and are described in detail in \citet{holtzman1995} 
 
\begin{figure} 
\epsscale{1.0} 
\plottwo{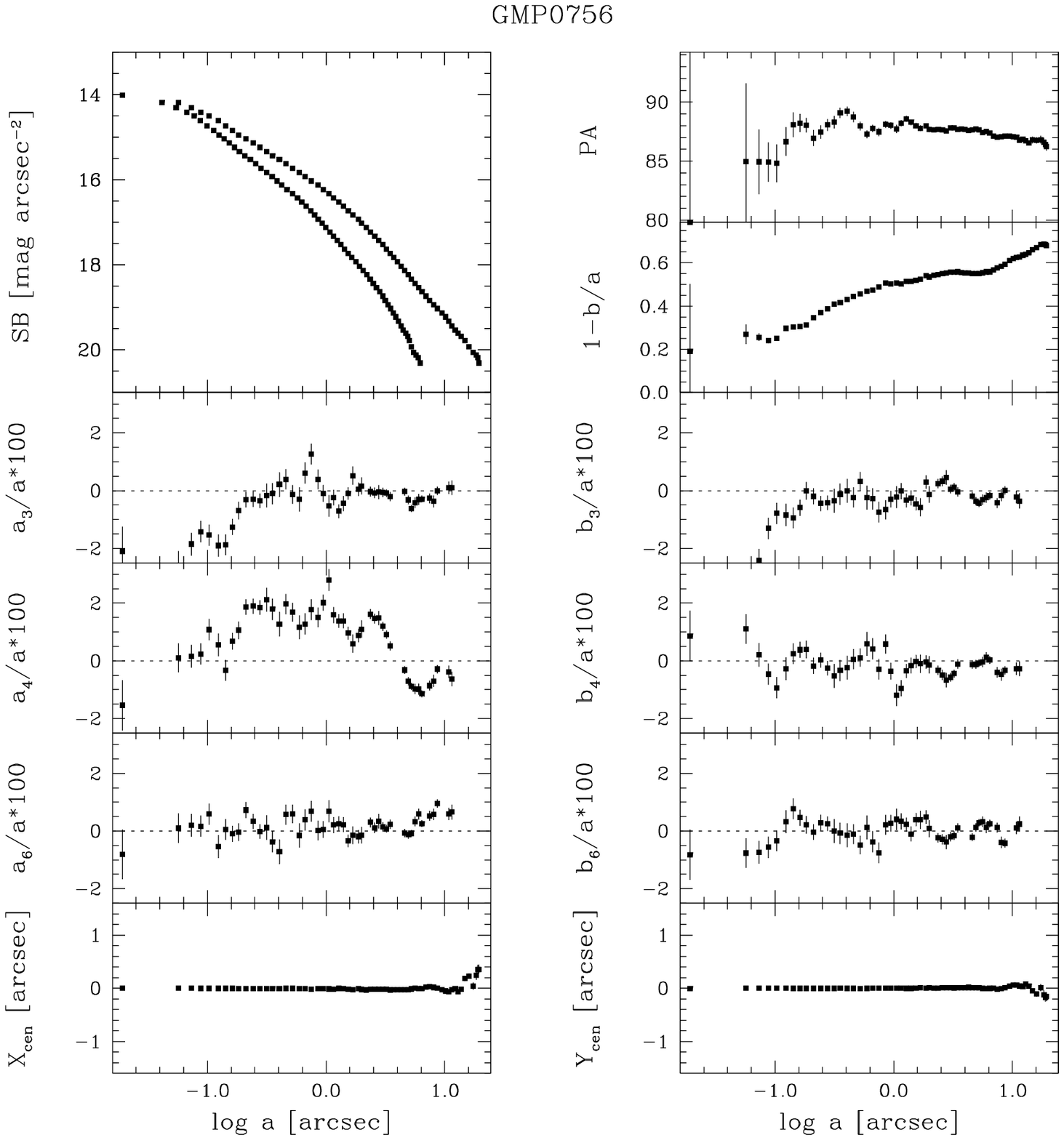}{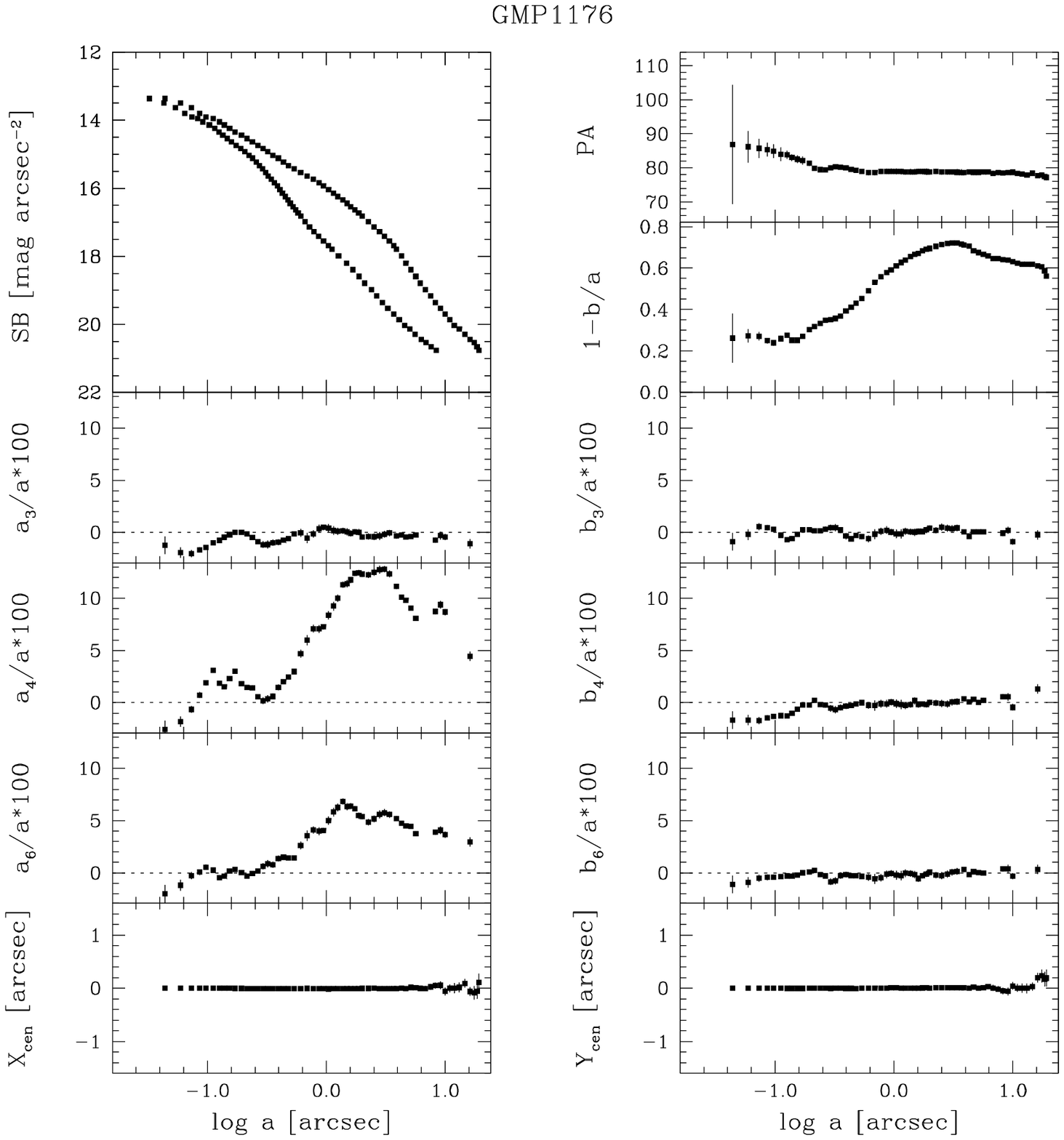}\\ 
\plottwo{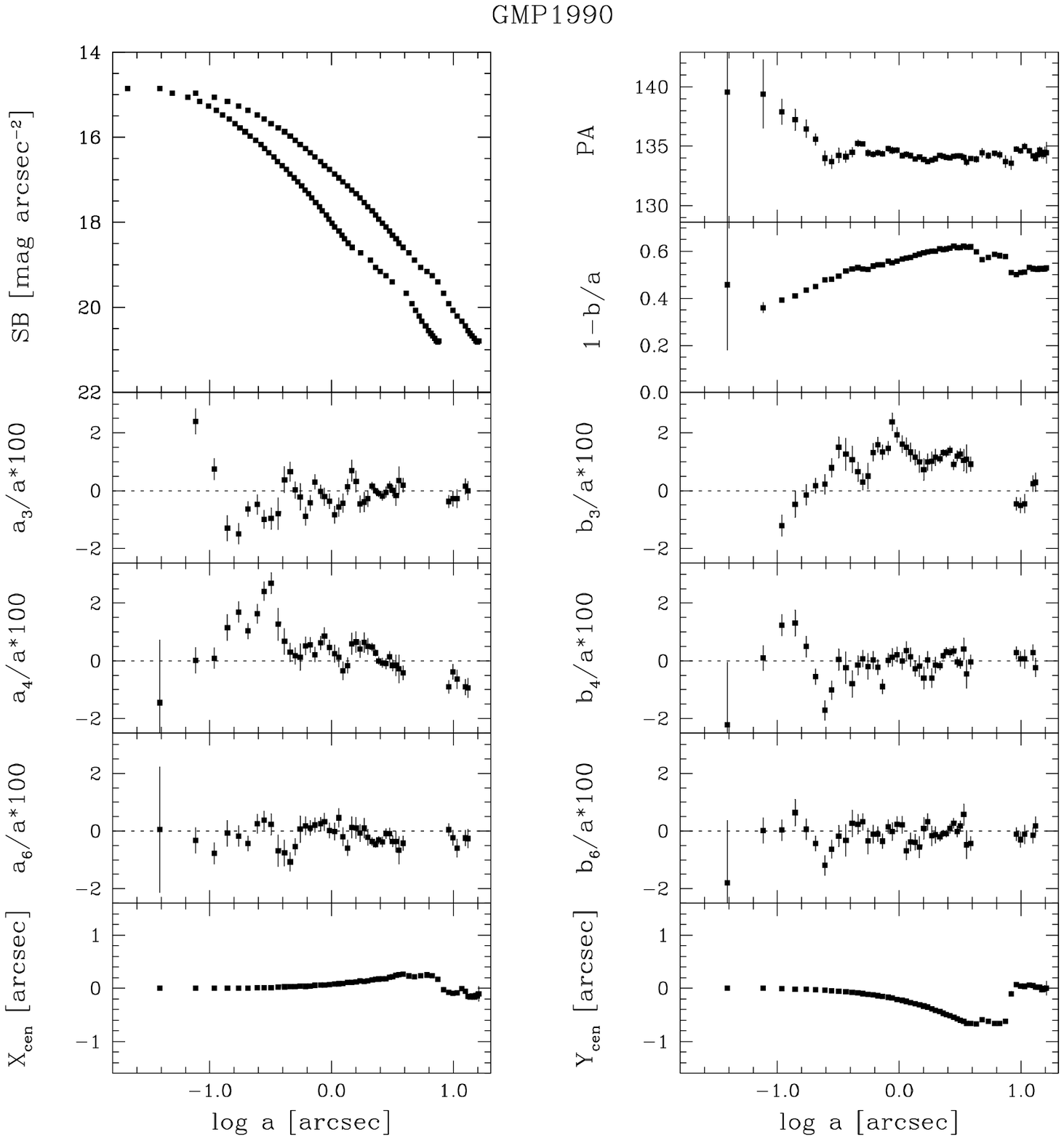}{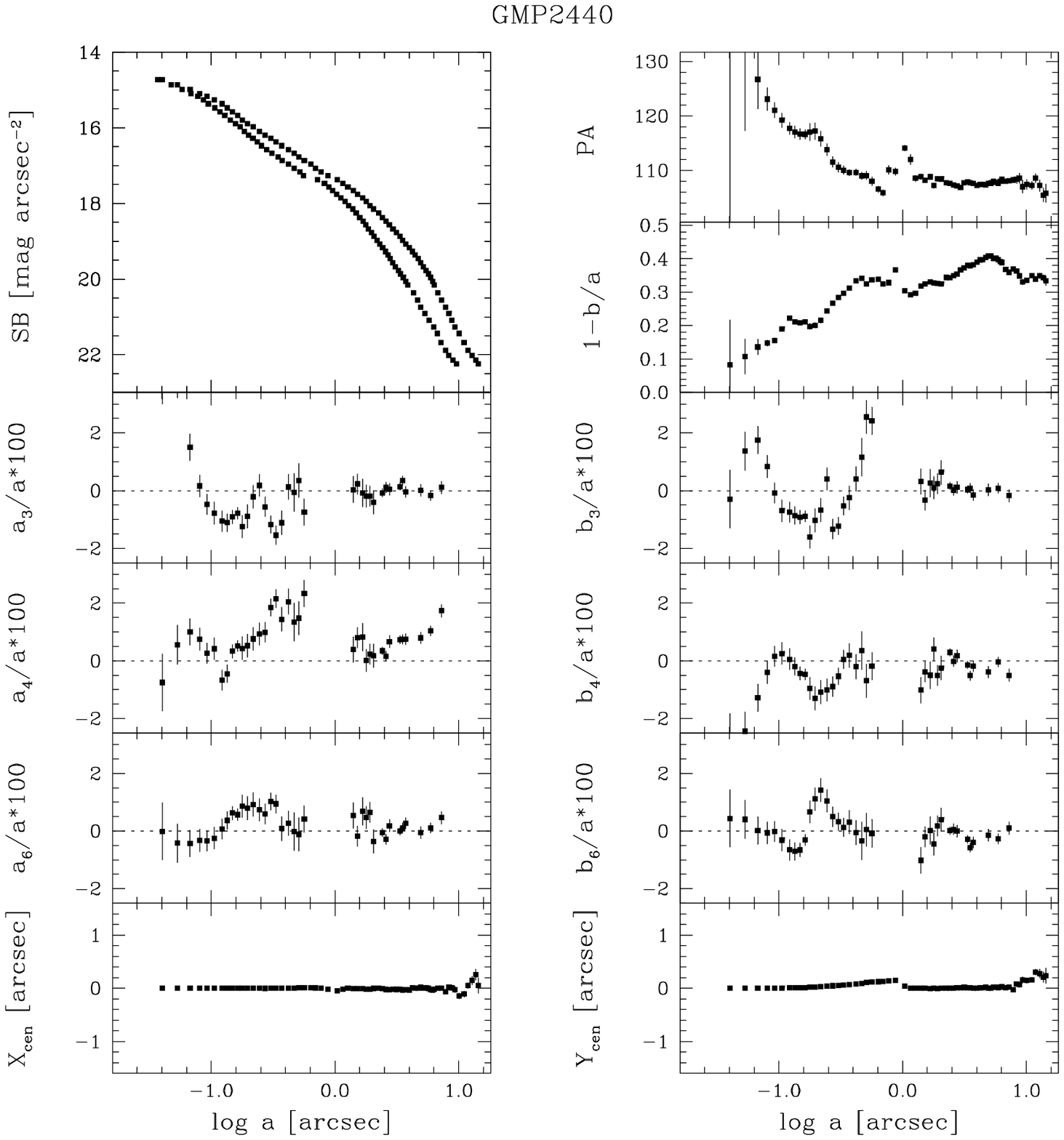} 
\caption{Isophotal parameters of the four galaxies recently 
  observed with HST/WFPC2 as a function of the logarithm of the
  semi-major axis distance in arcsec.
  The radial profiles of $R-$band surface brightness, third, fourth, 
  and sixth cosine Fourier coefficients ($a_3$, $a_4$, and $a_6$), and 
  $x-$coordinate of the center ($X_{\rm cen}$) are plotted in the left 
  panels (from top to bottom). The surface brightness is shown along 
  the major (upper profile) and minor (lower profile) axis. 
  The radial profiles of position angle (PA), ellipticity ($1-b/a$) 
  third, fourth, and sixth sine Fourier coefficients ($b_3$, $b_4$, 
  and $b_6$), and $y-$coordinate of the center ($Y_{\rm cen}$) are 
  plotted in the left panels (from top to bottom).} 
\label{fig:photometry} 
\end{figure}

The isophotal profiles of the galaxies were analyzed by fitting the 
isophotes with ellipses following the prescriptions by 
\citet{bender1987}. 
Foreground stars, cosmic rays, bad pixels, etc., were masked before 
fitting. Moreover, the centers of ellipses were allowed to vary.  As a 
first step, the skybackground was measured at the outer edges of the 
chip. Its final value was determined through the matching of the 
ground-based photometry (see below). 
A prominent dust lane present at the center of GMP~2440 was masked 
before performing the isophote analysis. Nevertheless deviations from 
axisymmetric isophotes are present at the level of $2\%$. A dust lane 
is also visible near the center of GMP~1990, causing the center 
coordinates to drift along the minor axis away from the dust lane and 
deviations from axisymmetric isophotes at the level of $2\%$.  Since 
masking the region affected by dust does not eliminate the drift or 
change the surface photometry, we give in Figure 
\ref{fig:photometry} and Table~\ref{tab:photometry} 
the results of the isophotal shape analysis obtained without masking 
the dusty regions. 
The surface photometry was calibrated in the $R$ band by matching the 
profiles given by \citet{jorgensen1995} for GMP~2440 and by Paper I 
for the other three objects. The matching of the surface brightness 
was performed between 5 and 15 arcsec by determinig the zero-point and 
sky value that minimize the square surface brightness flux 
differences. It gives rms of 0.01 mag for GMP~0756, 0.066 mag for 
GMP~1990, 0.08 mag for GMP~1176 and 0.038 mag for GMP~2440. The 
position angles match within 1$^\circ$, ellipticities within less than 
0.02. 
The radial profiles of the azimuthally averaged surface brightness,
ellipticity, position angle, center coordinates and third, fourth, and
sixth cosine ($a_3$, $a_4$, and $a_6$) and sine ($b_3$, $b_4$, and
$b_6$) Fourier coefficients are presented in Figure
\ref{fig:photometry} and Table~\ref{tab:photometry}.
 
\begin{figure} 
\epsscale{.80} 
\plotone{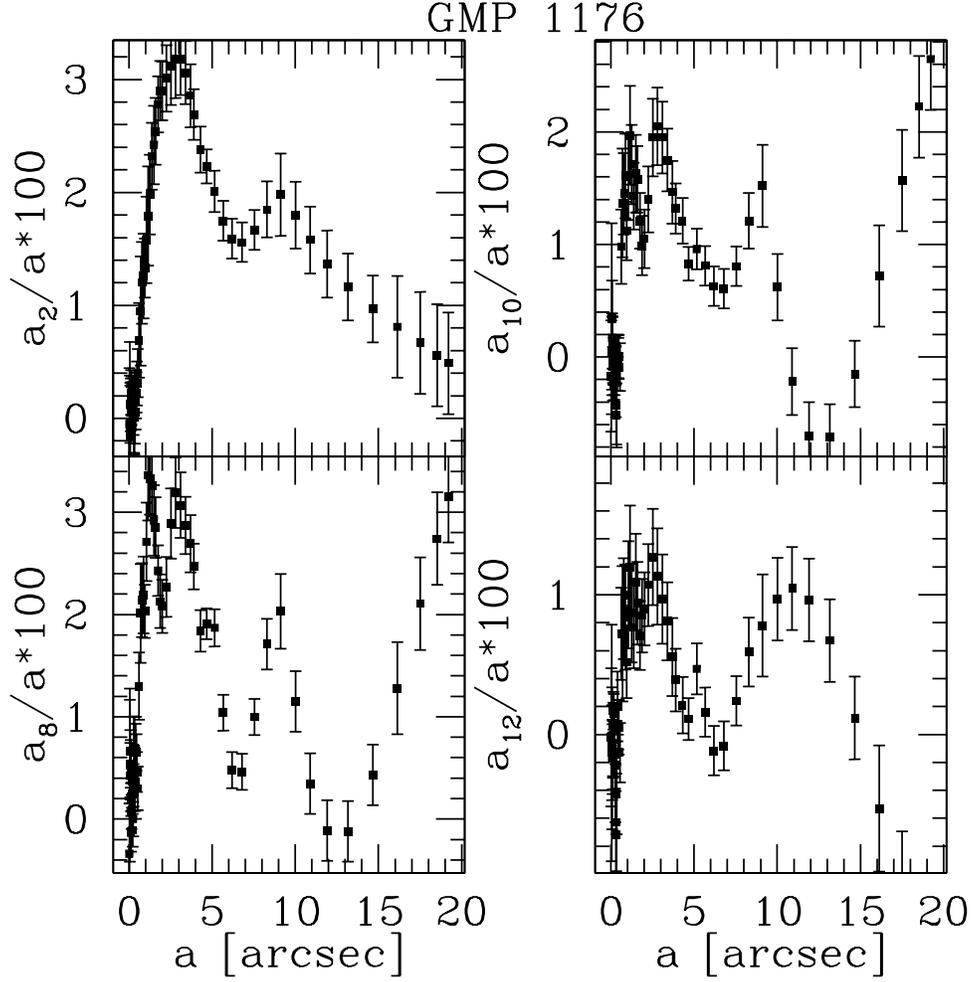} 
\caption{The radial profiles of the $R-$band second, eighth, tenth and  
  twelfth cosine Fourier coefficient ($a_2$, $a_8$, $a_{10}$, and $a_{12}$)  
  as a function of the semi-major axis distance in arcsec for the galaxy  
  GMP~1176.} 
\label{fig:phot1176} 
\end{figure}

GMP~1176 has extremely high ellipticity ($\le 0.75$) and diskiness 
$a_4$ exceeding 10\%. In this case the isophote fit program of 
\citet{bender1987} delivers a non-zero $a_2$ coefficient and a 
satisfactory description of the isophotes requires orders as high as 
$a_{12}$. The corresponding coefficient profiles are shown in Figure 
\ref{fig:phot1176} and listed in Table 
\ref{tab:phot1176}. GMP~0756 and GMP~1990 have also high ellipticity 
($\le 0.65$) but only mild ($\le 2$\%) diskiness.  GMP~2440 is 
slightly disky.  Overall, the high ellipticities and the morphology of 
the dust distribution indicate that all the four objects are very 
nearly edge-on.

\section{Spectroscopy} 
\label{sec:spectroscopy} 
 
\subsection{Galaxy sample}  
\label{sec:sample} 
 
All the observed galaxies (Tab. \ref{tab:log}) belongs to sample of 35
E and S0 galaxies of the Coma cluster studied in Paper I. For details
about their morphological classification and relevant photometric
properties (i.e., total magnitude, effective radius, mean surface
brightness within effective radius, ellipticity at effective radius,
and luminosity weighted $a_4$ parameter) the reader is referred to
that paper.

\subsection{Long-slit spectroscopy} 
\label{sec:longslit} 
 
Long-slit spectroscopic data of the sample galaxies were obtained with
9.2-m Hobby-Eberly Telescope (HET) at McDonald Observatory, Texas in
queuing mode \citep{shetrone2007} and with the 2.4-m Hiltner telescope
of the MDM Observatory at Kitt Peak, Arizona in visitor mode 
during different runs between 2002 and 2005.
Details of the instrumental set-up of the observations carried out on 
26 February -- 10 July 2002 (run 1), 11--15 April 2003 (run 2), 25--30 
May 2004 (run 3), 10 May 2005 (run 4), 07--10 April 2005 (run 
5) are given in Table~\ref{tab:setup}. 
 
Minor-axis spectra were obtained for all the sample galaxies, except 
for GMP~5975 which was observed only along a diagonal axis. The 
elliptical galaxies GMP~0144, GMP~2440, and GMP~4928 were observed 
along a diagonal axis too. Offset spectra with the slit parallel to 
the major axis were obtained only for the lenticular galaxies 
GMP~2417, GMP~3414, and GMP~5568. 
To perform a consistency check between measurements of stellar 
kinematics and line strength indices of different observing runs, 
diagonal-axis spectra of GMP~4928 were obtained in both run 2 and 3 
and minor-axis spectra of GMP~5568 were taken in both run 2 and 
5. In addition, spectra along the minor axis of GMP~0144 were taken to 
be compared with measurements of Paper II. 
Typical integration time of galaxy spectra was 3600 s. Total 
integration times and slit position angle of the galaxy spectra as 
well as the log of the spectroscopic observations are given in Table 
\ref{tab:log}. 
At the beginning of each exposure the galaxy was centered on the slit
either using acquisition images with HET or the guiding camera with
the 2.4-m Hiltner telescope which looks onto the slit.
 
In each run several spectra of giant stars with spectral type ranging
from late-G to early-K were obtained to be used as template in
measuring stellar kinematics and line strength indices. The template
stars were selected from \citet{faber1985} and
\citet{gonzales1993}. Additionally, we observed at least one flux
standard star per night to calibrate the flux of the spectra before
line indices were measured. Spectra of the comparison arc lamp were
taken before and/or after object exposures (MDM) or at the end of the
night (HET) to allow an accurate wavelength calibration.
 
All the spectra were bias subtracted, flatfield corrected, cleaned of
cosmic rays, corrected for bad columns and wavelength calibrated using
standard {\tt IRAF}\footnote{{\tt IRAF} is distributed by NOAO, which
is operated by AURA Inc., under contract with the National Science
Foundation.} routines.
The flatfield correction was performed by means of quartz lamp spectra 
in runs 1 and 4, of both quartz lamp and twilight sky spectra in runs 
2,3, and 5. They were normalized and divided into all the spectra, to 
correct for pixel-to-pixel sensitivity variations and large-scale 
illumination patterns due to slit vignetting. Cosmic rays were 
identified and corrected by interpolating over as in 
\citet{bender1994}. The residual cosmic rays were corrected by 
manually editing the spectra. 
Each spectrum was rebinned using the wavelength solution obtained from
the corresponding arc-lamp spectrum. We checked that the wavelength
rebinning had been done properly by measuring the difference between
the measured and predicted wavelengths for the emission lines in the
comparion arc lamp spectra. The rms of the wavelength solution is
$\sim0.10$ \AA\ in run 1, $\sim0.06$ \AA\ in runs 2, 3, and 5, and
$\sim0.36$ \AA\ in run 4. They correspond to $\sim6$, 3, and 21 \kms\
at 5170 \AA\ (i.e., the wavelength of the \MgI\ absorption triplet),
respectively. Systematic errors of the absolute wavelength calibration
($\leq10$ \kms ) were estimated by measuring the difference between
the measured and predicted wavelengths \citet{osterbrock1996} for the
brightest night-sky emission lines in the observed spectral range.
The instrumental resolution in each run was derived as the mean of the 
Gaussian FWHMs measured for a number of unblended arc-lamp lines which 
were distributed over the whole spectral range of a 
wavelength-calibrated spectrum. The mean FWHM of the arc-lamp lines 
and the corresponding instrumental resolution derived at 5170 \AA\ is 
given in Table \ref{tab:setup}. 
In galaxy and stellar spectra the contribution of the sky was
determined by interpolating along the outermost 10--30 arcsec at the
two edges of the slit, where the galaxy or stellar light was
negligible, and then subtracted. A sky subtraction better than $1\%$
was achieved.  All the spectra were corrected for CCD misalignment
following \citet{bender1994}. The spectra obtained for the same galaxy
along the same axis were coadded using the center of the stellar
continuum as reference. This allowed improving the $S/N$ of the
resulting two-dimensional spectrum. A one-dimensional spectrum was
obtained for each kinematical template star as well as for each flux
standard star. The spectra of the kinematical Lick-system templates
were deredshifted to laboratory wavelengths.
The FWHM of the point spread function (PSF) due to seeing and
instrumental set-up is given in Tab. \ref{tab:log} for the coadded
galaxy spectra. It was estimated by comparing the surface-brightness
radial profile obtained from each spectrum with the corresponding one
extracted from the HST/WFPC2 $R-$band image of the galaxy. All the
profiles are presented in \citet{jthomas2007}. Slit width and
orientation and WFPC2 PSF were taken into account. The low $S/N$ in
the outskirts of the WFPC2 image of GMP~5568 did not allowed to
extract a reliable surface-brightness profile to be compared to that
of farthest offset spectrum the galaxy. The FWHM of this spectrum was
derived by interpolating the values of the spectra obtained before and
after it. The same is true for the spectra of GMP~3414 and GMP~4822,
for which no image was available in the HST Science Archive.

\section{Results} 
\label{sec:results} 
 
\subsection{Stellar kinematics}  
\label{sec:kinematics}

The stellar kinematics was measured from the galaxy absorption 
features present in the wavelength range and centered on the \MgI\ 
line triplet ($\lambda\lambda\,5164,5173,5184$ \AA) by applying the 
Fourier Correlation Quotient method \citep{bender1990} as done by 
\citet{bender1994}. 
  
\begin{figure} 
\epsscale{.80} 
\plotone{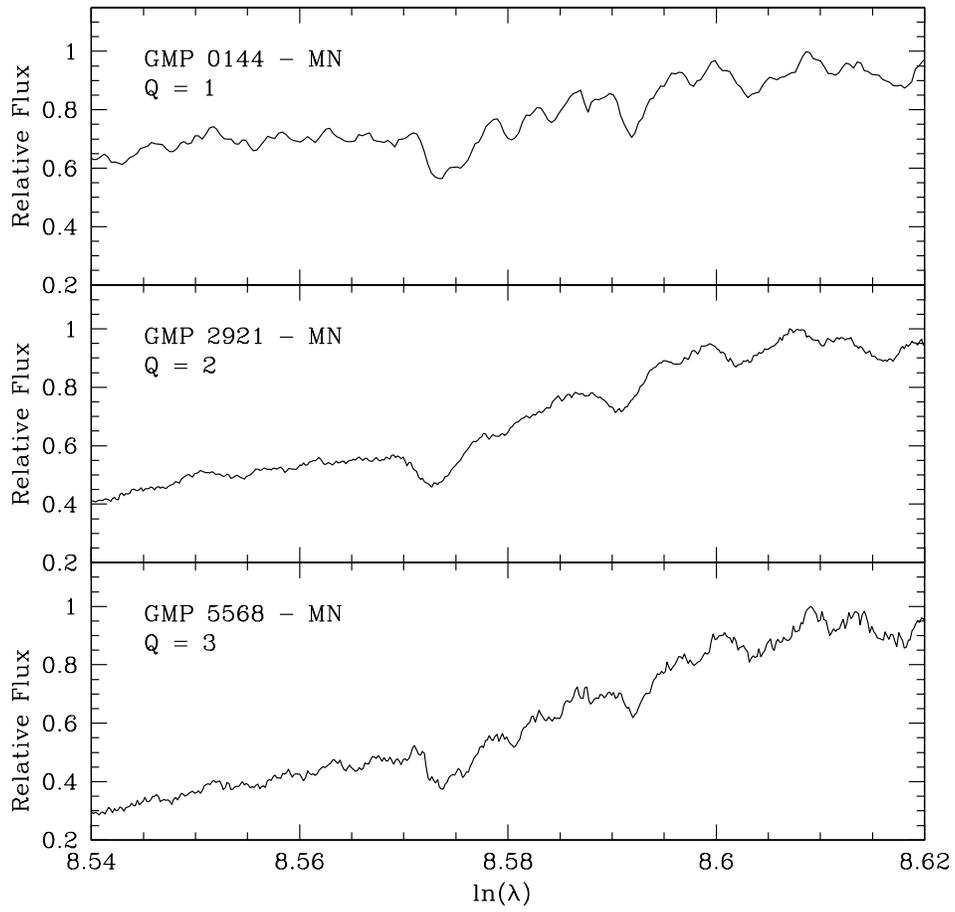} 
\caption{Example of central spectra covering the range  
  of quality classes. Relative fluxes have false zero points for 
  viewing convenience.} 
\label{fig:quality} 
\end{figure}

The spectra were rebinned along the dispersion direction to natural 
logarithmic scale, and along the spatial direction to obtain a nearly 
constant $S/N \geq 20$ per resolution element. In few spectra the 
$S/N$ decreases to $\sim10$ at the outermost radii. The galaxy 
continuum was removed row-by-row by fitting a fourth to sixth order 
polynomial. 
The quality of the final spectrum depends on resulting $S/N$. In Fig. 
\ref{fig:quality} we show example of central spectra covering the 3 
quality classes listed in Table \ref{tab:log}. The quality 
parameter is 1 for $S/N\geq100$, 2 for $50 \leq S/N < 100$, 3 for $20 
\leq S/N \leq 50$. 
 
To measure the stellar kinematics of the sample galaxies we adopted 
HR~6817 (K1III) as kinematical template for run 1, HR 6018 (K1III) for 
run 2, and HR 3427 (K0III) for runs 3--5. 
We derived for each galaxy spectrum the line-of-sight velocity
distribution (LOSVD) along the slit and measured its moments, namely
the line-of-sight radial velocity $v$, velocity dispersion $\sigma$
and values of the coefficients $H_3$ and $H_4$.  At each radius, they
have been derived by fitting the LOSVD with a Gaussian plus third- and
fourth-order Gauss-Hermite polynomials ${\cal H}_3$ and ${\cal H}_4$,
which describe the asymmetric and symmetric deviations of the LOSVD
from a pure Gaussian profile \citep{vandermarel1993,gerhard1993}.
Errors on the LOSVD moments were derived from photon statistics and 
CCD read-out noise, calibrating them by Monte Carlo simulations as 
done by \citet{bender1994}. In general, errors are in the range of 
3--10 \kms\ for $v$ and $\sigma$, and of 0.01--0.04 for $H_3$ and 
$H_4$, becoming larger in the outer parts of some galaxies where for 
$10 \la S/N < 20$. The largest errors are observed for offset spectra 
of GMP~2417, GMP~3414, and GMP~5568 with $S/N \approx 20$. 
These errors do not take into account possible systematic effects due 
to template mismatch or the presence of dust and/or faint 
emission. The measured stellar kinematics is reported in Table 
\ref{tab:kinematics} and plotted in Fig. \ref{fig:kinematics}, where 
profiles folded with respect to, and velocities relative to the galaxy 
centers are given. 
 
Fig. \ref{fig:comparison} shows the comparison between the 
measurements of $v$, $\sigma$, $H_3$, and $H_4$ along the minor axis 
of GMP~0144 obtained here and measurements of Paper II. There is a 
significant difference in velocity dispersions within 3 arcsec, that 
can be only partially attributed to the different instrumental set-up 
and seeing of the observing runs. The same is true for the velocity 
dispersion observed along the diagonal (Fig. \ref{fig:kinematics}) and 
major axis (Paper I). 
A likely explanation for the mismatch is that GMP~0144 deviates 
significantly from axisymmetry near its center and hosts a dynamically 
hot, decoupled component \citep{jthomas2006,jthomas2007}. Further 
observational support for this comes from the relatively large 
side-to-side asymmetries in all the measured slits and the strong 
isophotal twist ($\sim30^\circ$) measured in the very center from 
HST/WFPC2 imaging (Paper I). 
 
The multiple observations of GMP~4928 and GMP~5568 agree within the 
errors.

\subsection{Line strength indices}  
  
We measured the Mg, Fe, and \Hb\ line strength indices following 
\citet{faber1985} and \citet{worthey1994} from flux calibrated 
spectra, as done in Paper I and II.  Spectra were rebinned in 
dispersion direction as well as in radial direction as before. We 
indicate the average Iron index with \Fe $\rm = 
(Fe_{5270}+Fe_{5335})/2$ \citep{gorgas1990} and the usual combined 
Magnesium-Iron index with $\rm [MgFe] = \sqrt{Mg{\it b} \langle Fe 
\rangle}$ \citep{gonzales1993}. We corrected all the measured indices 
for velocity dispersion broadening and calibrated our measurements to 
the Lick system using stars from \citet{worthey1994} 
(Fig.~\ref{fig:lick}). No zero-point correction was applied since data 
did not show offset with respect to the Lick system.  
No focus correction was applied because atmospheric seeing was the
dominant effect during observations (see \citealt{mehlert1998} for
details).
Errors on indices were derived from photon statistics and CCD read-out 
noise, and calibrated by means of Monte Carlo simulations. 
 
The strongest emission feature in the observed spectral range is the 
\OIII 5007\AA\ emission line. It falls on top the Fe$_{5015}$ Iron 
index which we neglected and it is connected to \Hb\ emission line 
\citep{osterbrock1989}. By adopting the index definition by  
\citet{gonzales1993} the equivalent width of the \OIII 5007\AA\  
emission line fell below the detection limit in all but few of the
spectra, namely those obtained along the minor axis of GMP~2921,
GMP~3329, and GMP~4822, and along the diagonal axis of GMP~2440 and
GMP~4928. Even in these spectra the \OIII 5007\AA\ equivalent width
was $\lesssim0.5$ \AA . This would correspond to a correction
$\lesssim0.3$ \AA\ for the equivalent width of \Hb . Since it is
typically smaller than the statistical errors on \Hb\ measurements, no
correction for emission was applied.
 
The measured values of \Hb , [MgFe], \Fe, \Mgb , and \Mgd\ are listed 
in Table \ref{tab:indices} and plotted in Fig. \ref{fig:kinematics}. 
Fig. \ref{fig:comparison} shows the comparison between the 
measurements of \Hb , [MgFe], \Fe, \Mgd , and \Mgb\ along the minor 
axis of GMP~0144 obtained here and measurements obtained in Paper 
II. The values derived from the two datasets are in agreement within 
the errors. 
Fig. \ref{fig:central} shows the central values of $\sigma$, 
\Hb , [MgFe], \Fe , \Mgb , and \Mgd\ obtained along the minor and 
diagonal axes of the sample galaxies as a weighted mean of the values 
available within an aperture of 2 arcsec and those obtained along the 
corresponding major axes. Most of the data are consistent within 
$3\sigma$ errors. No systematic effects are observed for the remaining 
ones.

\section{Conclusions} 
\label{sec:conclusions} 
 
New radially resolved spectroscopy of 12 E and S0 galaxies of the Coma
cluster was presented. The rotation curves, velocity dispersion
profiles and the $H_3$ and $H_4$ coefficients of the Hermite
decomposition of the line-of-sight velocity distribution were derived
along the minor axis, offset major axis and one diagonal
direction. Moreover, the line strength index profiles of Mg, Fe and
\Hb\ line indices were measured too.
In addition, the surface photometry of the central regions of a subsample 
of 4 galaxies recently obtained with HST/WFPC2 was presented. 
 
The data complement the existing set (Paper I, Paper II) and have a 
precision and radial extent sufficient to construct flattened and 
rotating dynamical models of the galaxies and study their radially 
resolved stellar populations. Other papers address these issues (Paper 
III, \citealt{jthomas2005,jthomas2007}).

\acknowledgments  
EMC acknowledge the Max-Planck-Institut f\"ur extraterrestrische 
Physik for hospitality while this paper was in progress. 
EMC receives support from the grant PRIN2005/32 by Istituto Nazionale
di Astrofisica (INAF) and from the grant CPDA068415/06 by the Padua
University.
This work was supported by the Sonderforschungsbereich 375
'Astro-Teilchenphysik' of the Deutsche Forschungsgemeinschaft.
Support for Program number HST-GO-10884.0-A was provided by NASA
through a grant from the Space Telescope Science Institute which is
operated by the Association of Universities for Research in Astronomy,
Incorporated, under NASA contract NAS5-26555.

%\clearpage 
 
%\clearpage 
 
%\clearpage 
 
%\clearpage 
 
\begin{figure} 
\epsscale{1.0} 
\plottwo{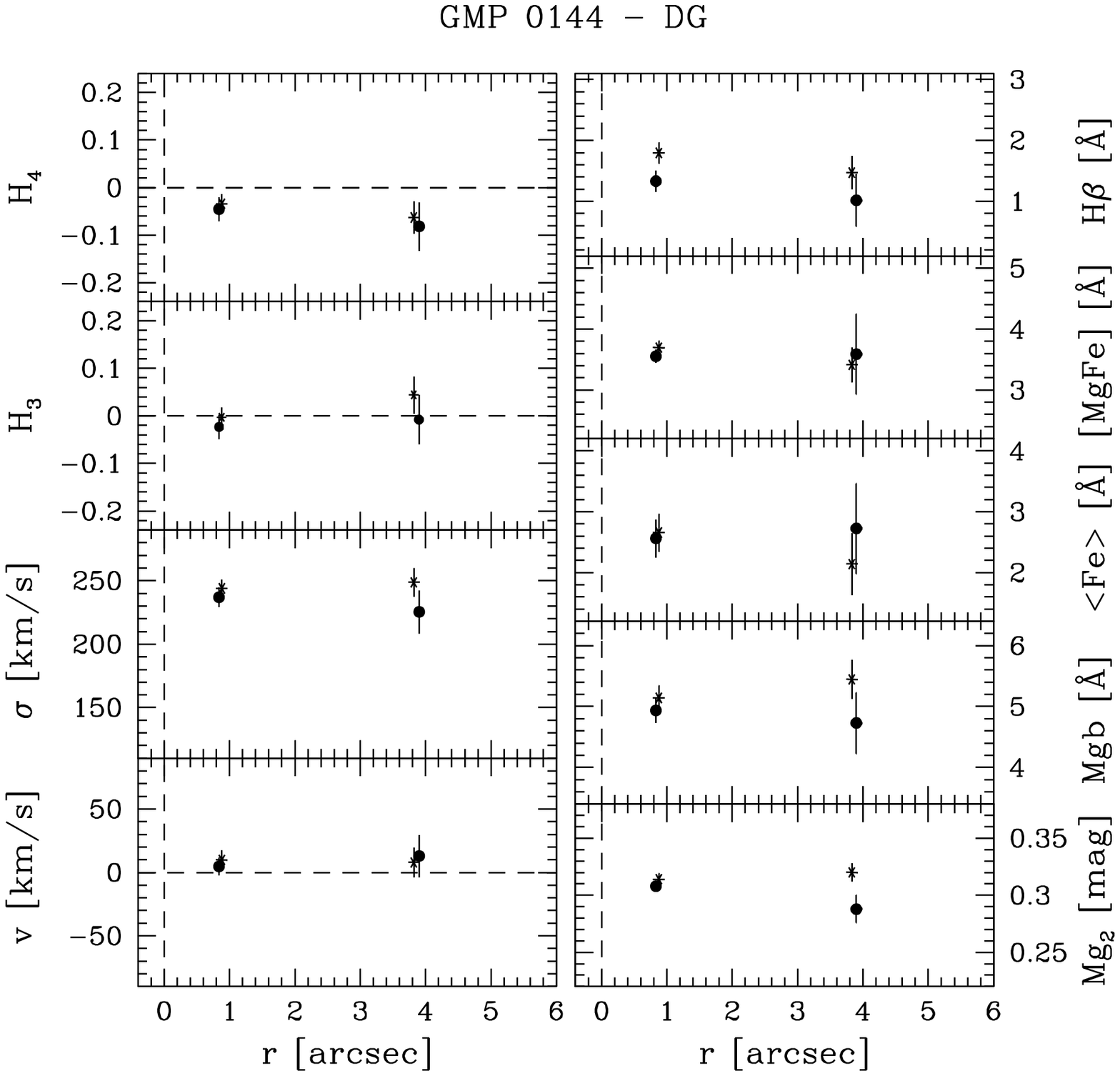}{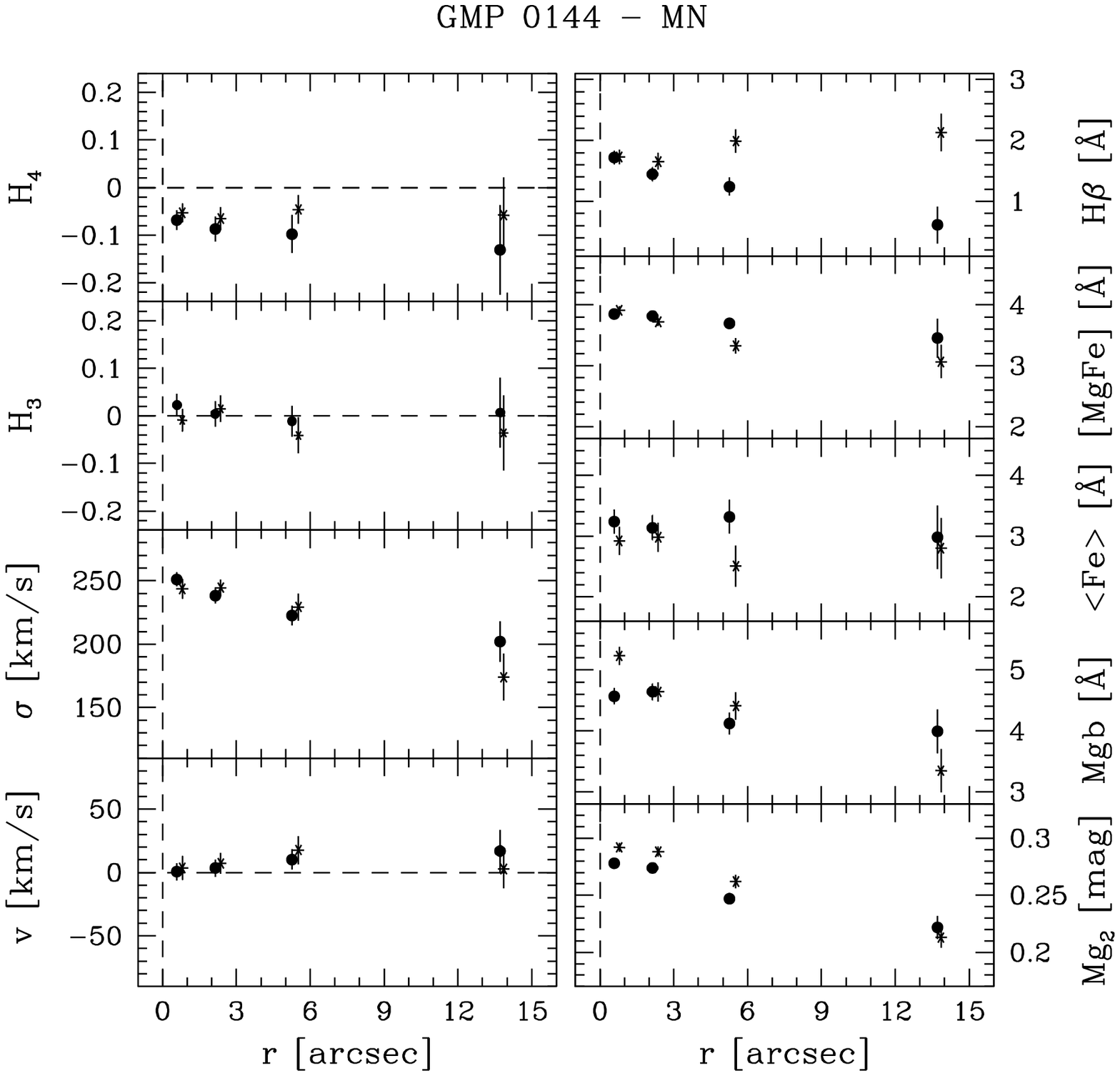}\\ 
\plottwo{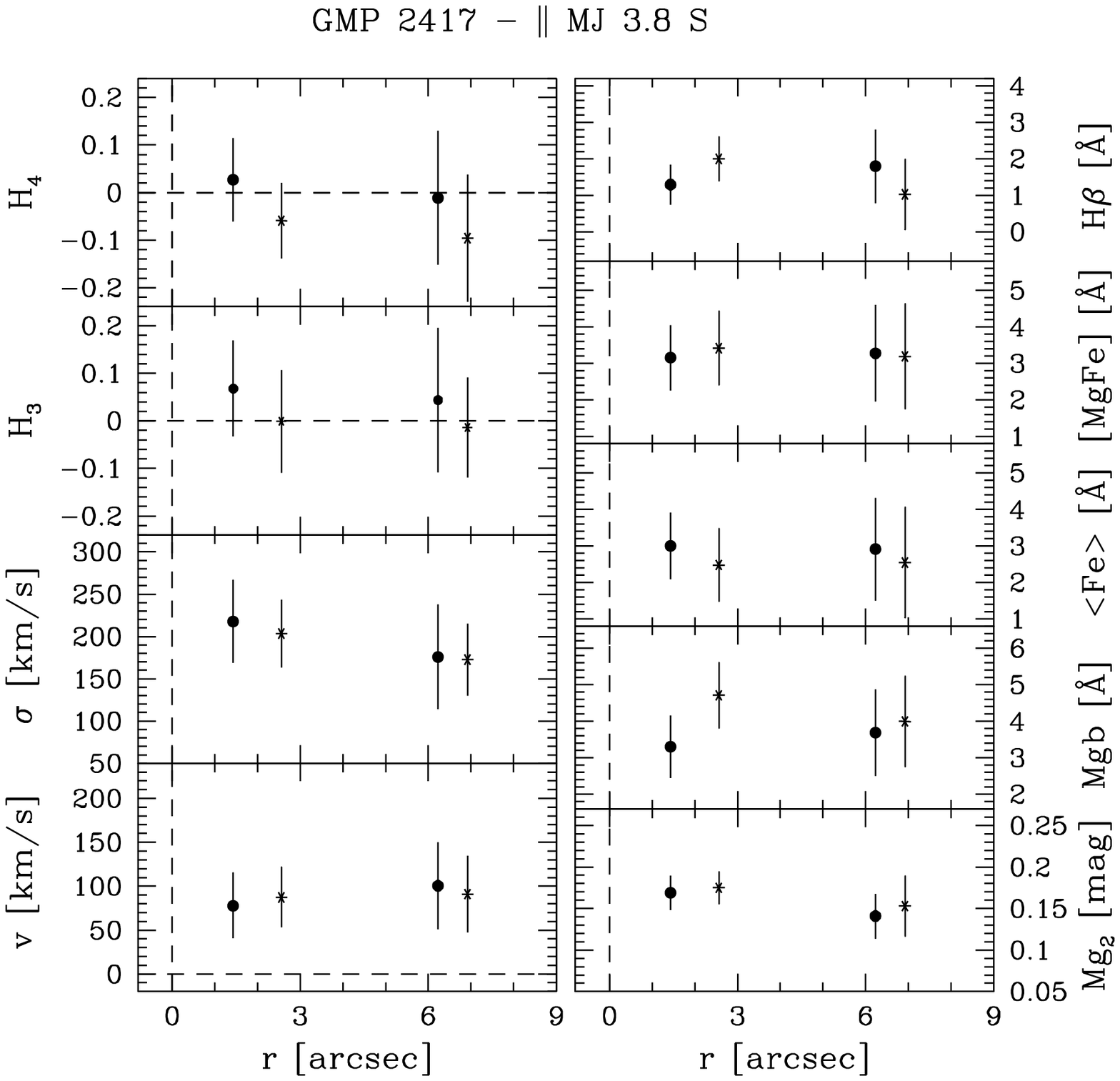}{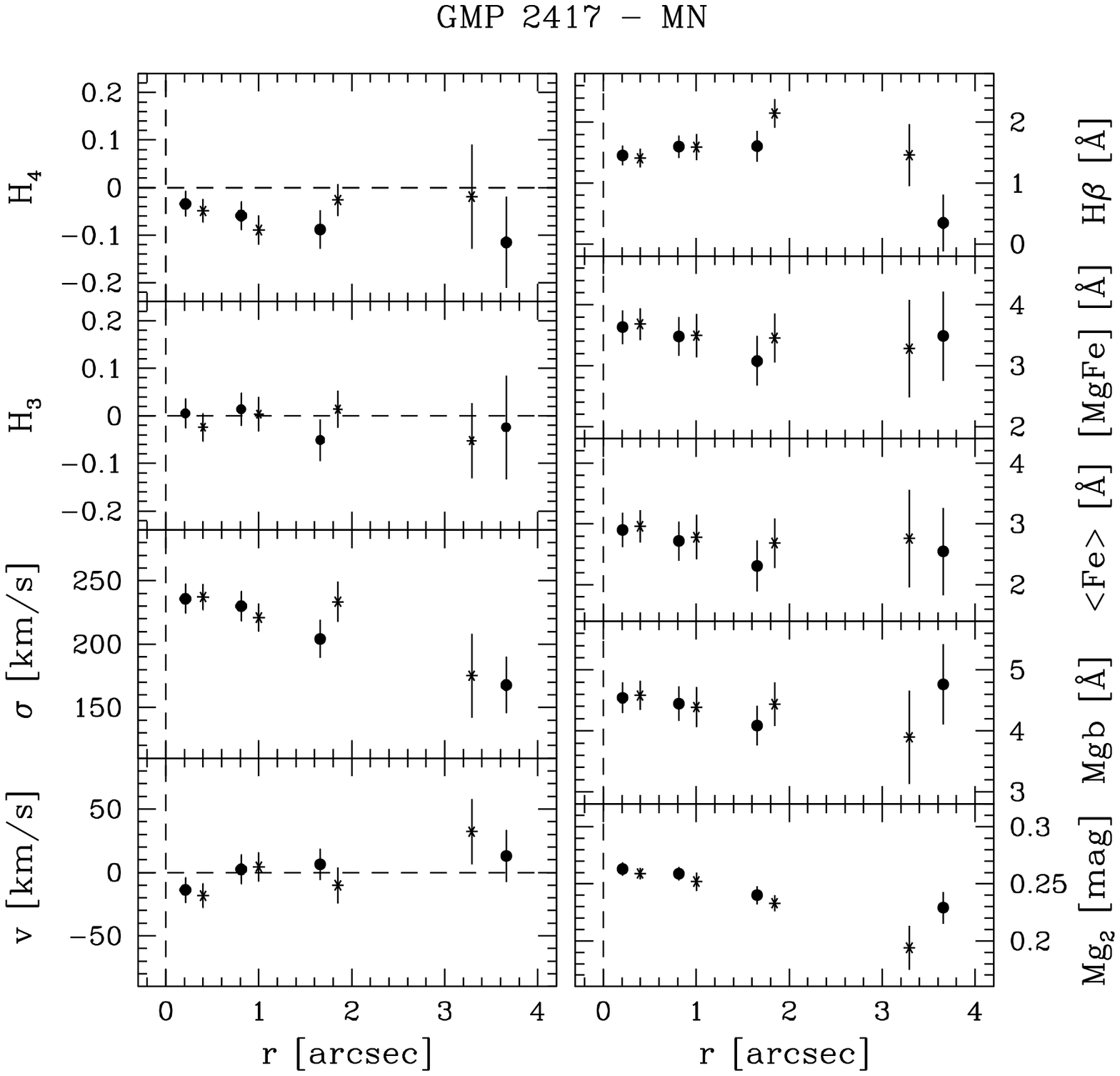} 
\caption{Kinematical parameters and line indices measured along the 
  observed axes of the sample galaxies. For each axis the curves are
  folded around the nucleus. Circles and asterisks (or squares) refer
  to data measured along the receding and approching side,
  respectively. The radial profiles of the line-of-sight velocity
  ($v$) after the subtraction of systemic velocity, velocity
  dispersion ($\sigma$), third ($H_3$) and fourth ($H_4$) order
  coefficient of the Gauss-Hermite decomposition of the LOSVD are
  shown in the left panels (from top to bottom).  The radial profiles
  of the line indices \Hb, [MgFe], \Fe , \Mgb , and \Mgd\ are plotted
  in the right panels (from top to bottom).
  For GMP~4928 open and filled symbols refer to data 
  measured from spectra obtained in run 2 and 3, respectively. 
  For GMP~5568 open and filled symbols refer to data 
  measured from spectra obtained in run 2 and 5, respectively.} 
\label{fig:kinematics}   
\end{figure} 
 
%\clearpage 
 
\addtocounter{figure}{-1} 
\begin{figure} 
\epsscale{1.0} 
\plottwo{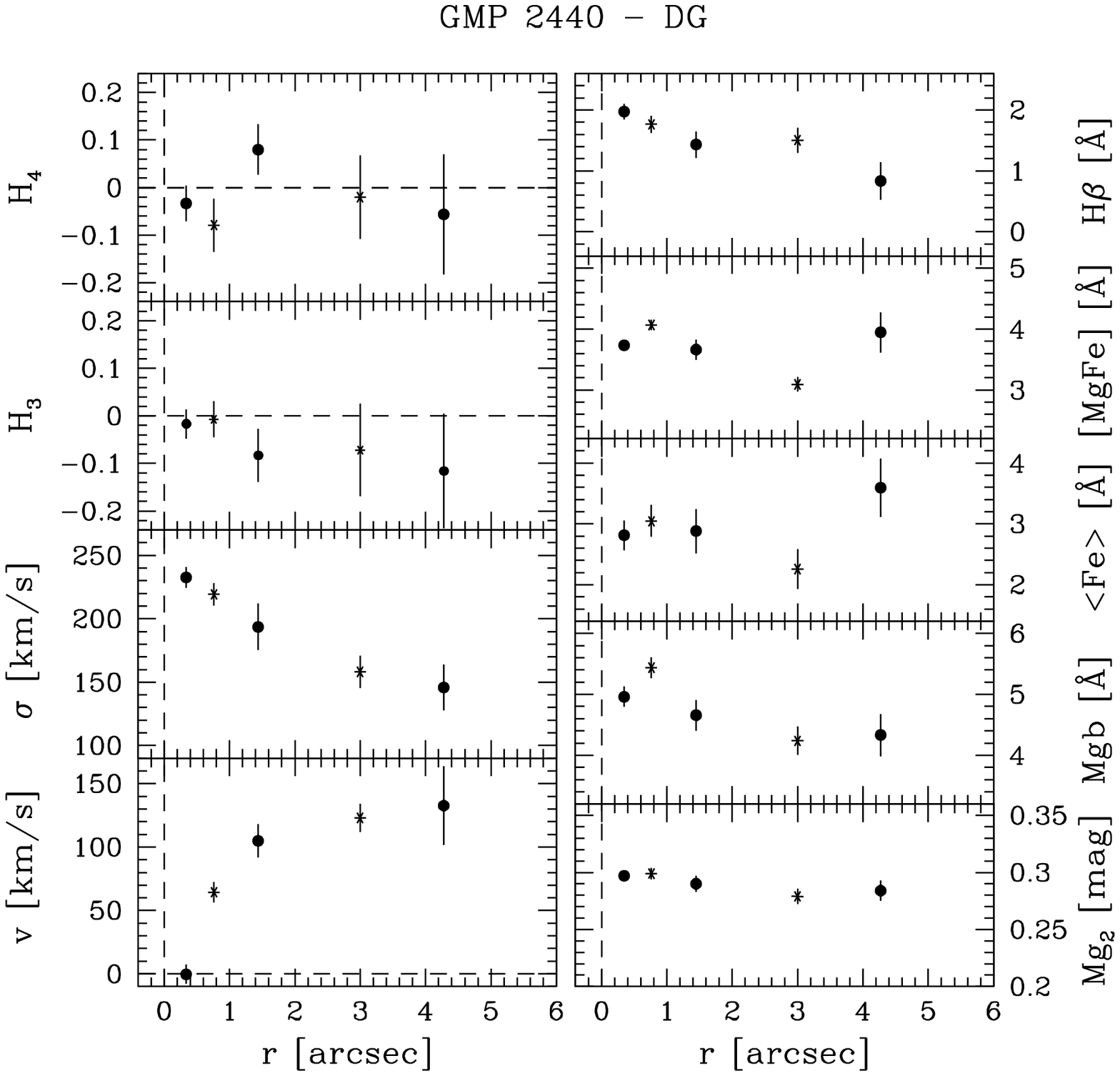}{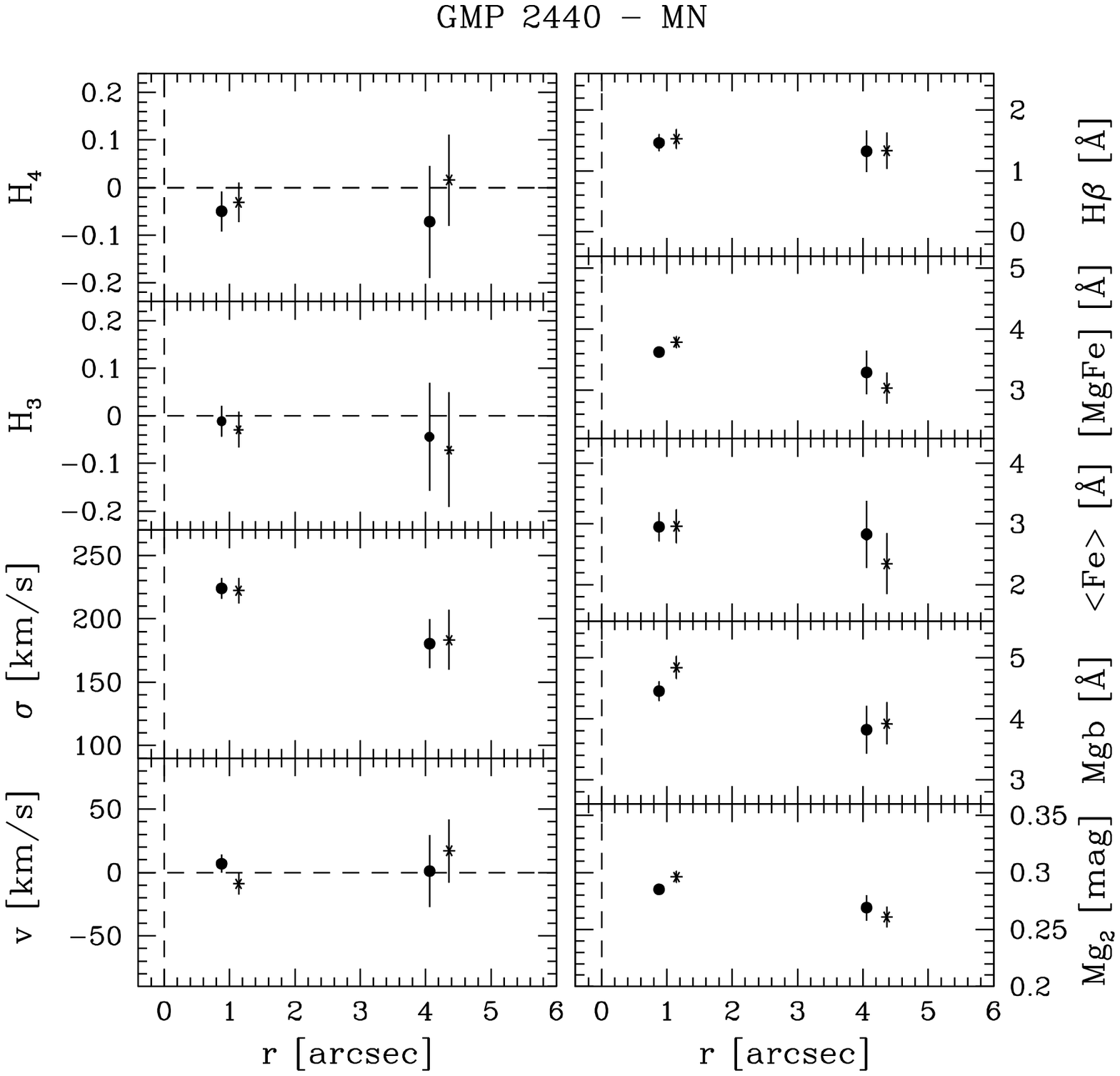}\\ 
\plottwo{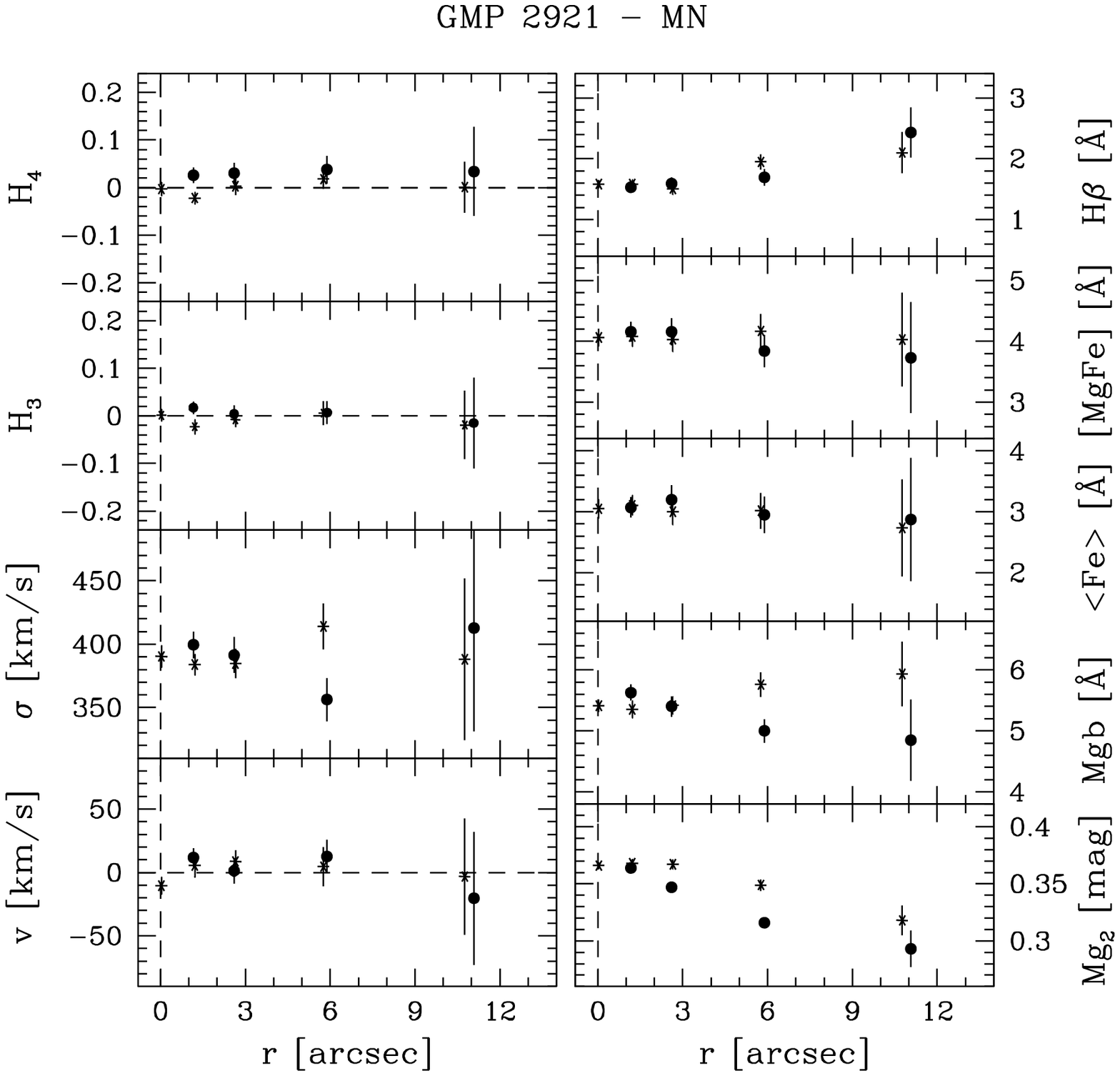}{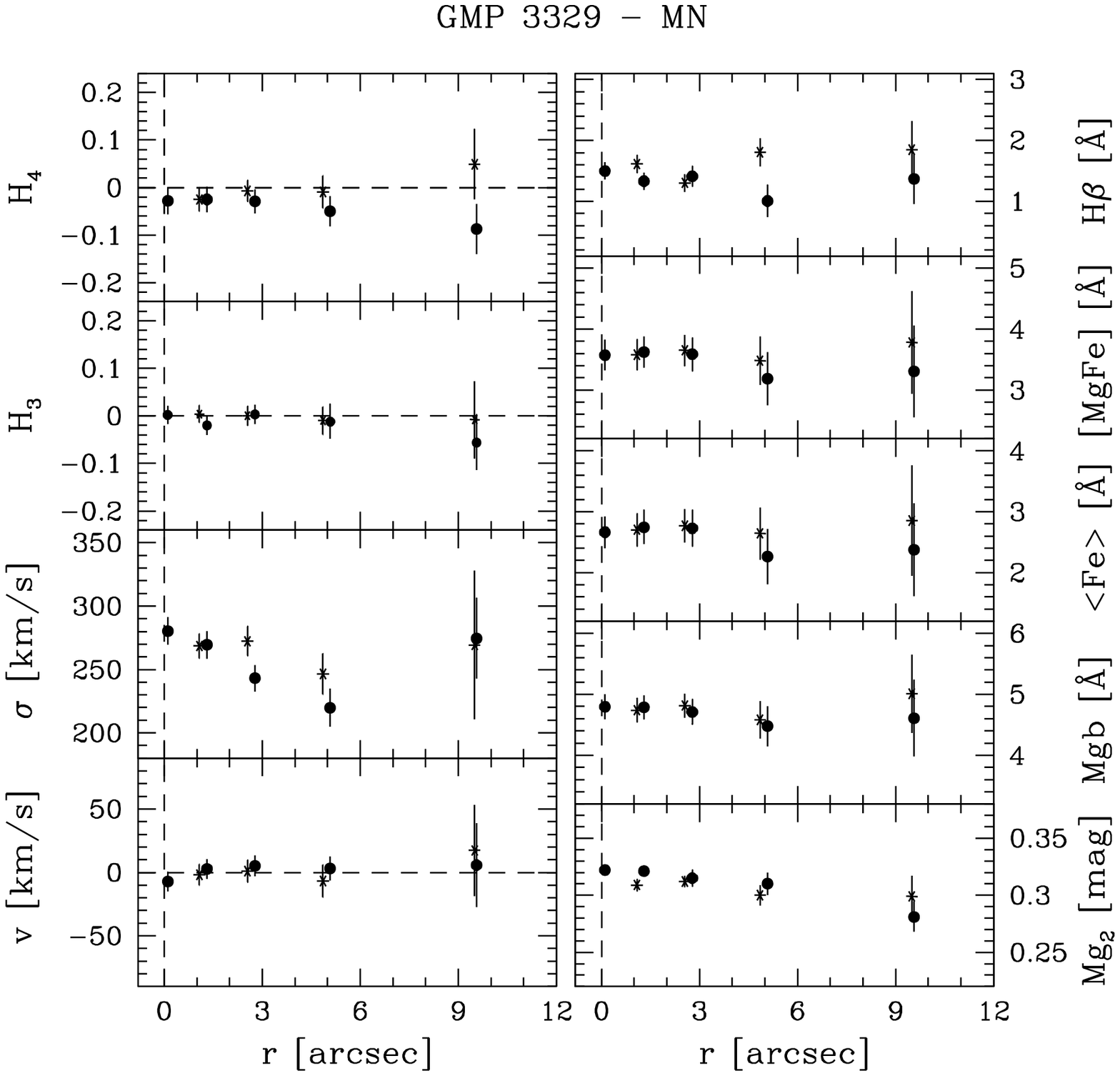}\\ 
\plottwo{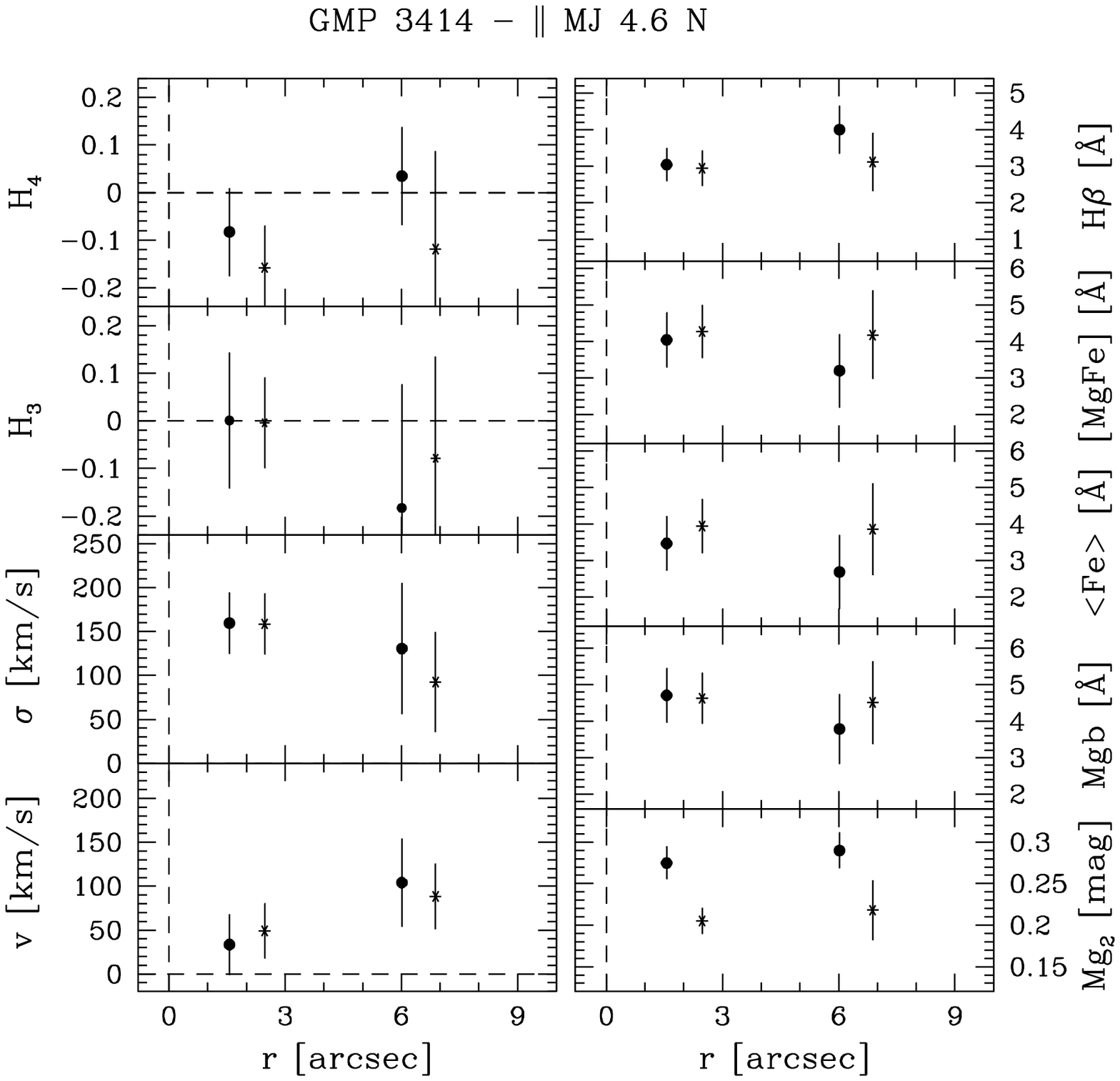}{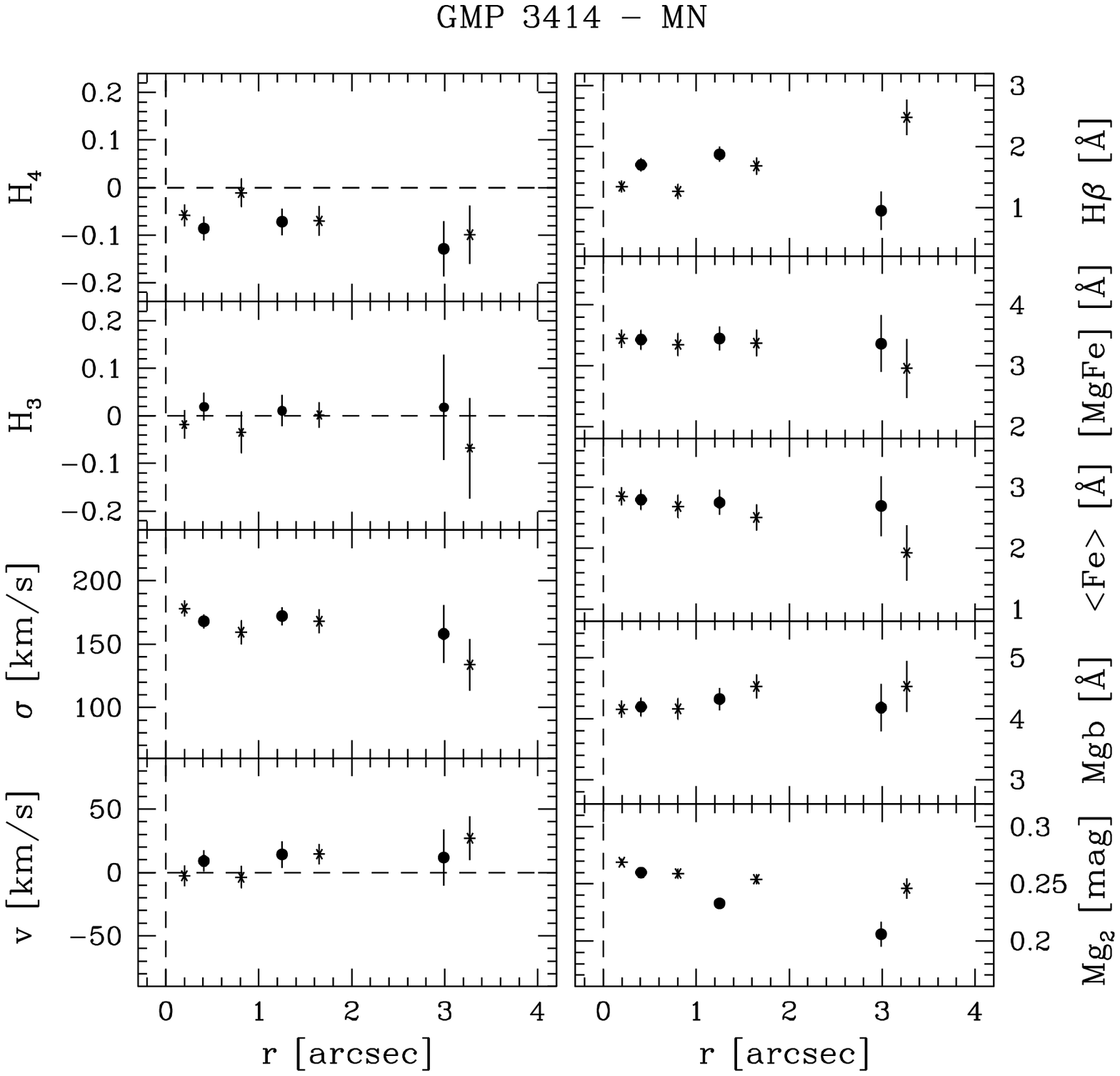} 
\caption{Continued.} 
%\label{fig:kinematics}   
\end{figure}

%\clearpage 
 
\addtocounter{figure}{-1} 
\begin{figure} 
\epsscale{1.0} 
\plottwo{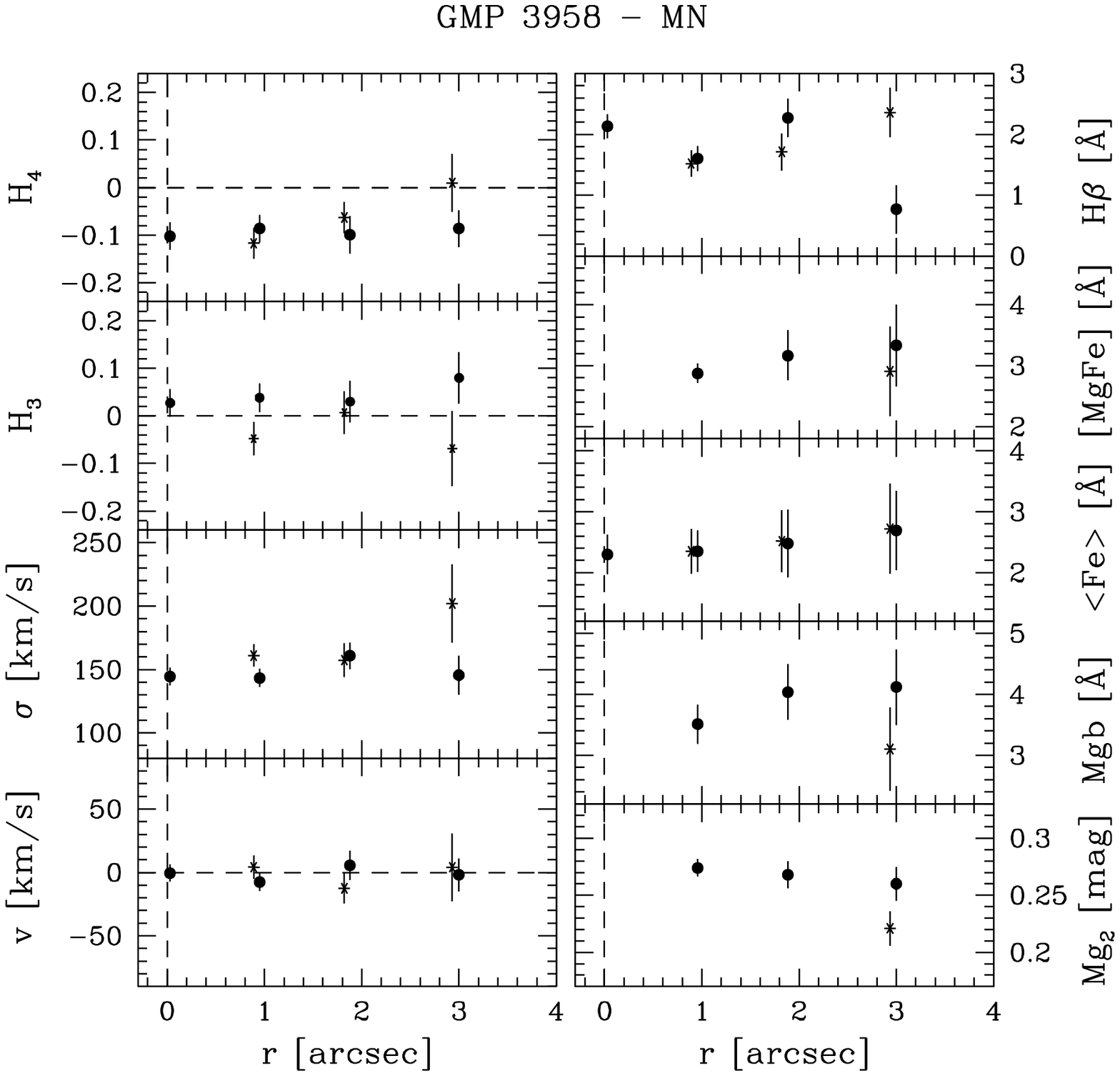}{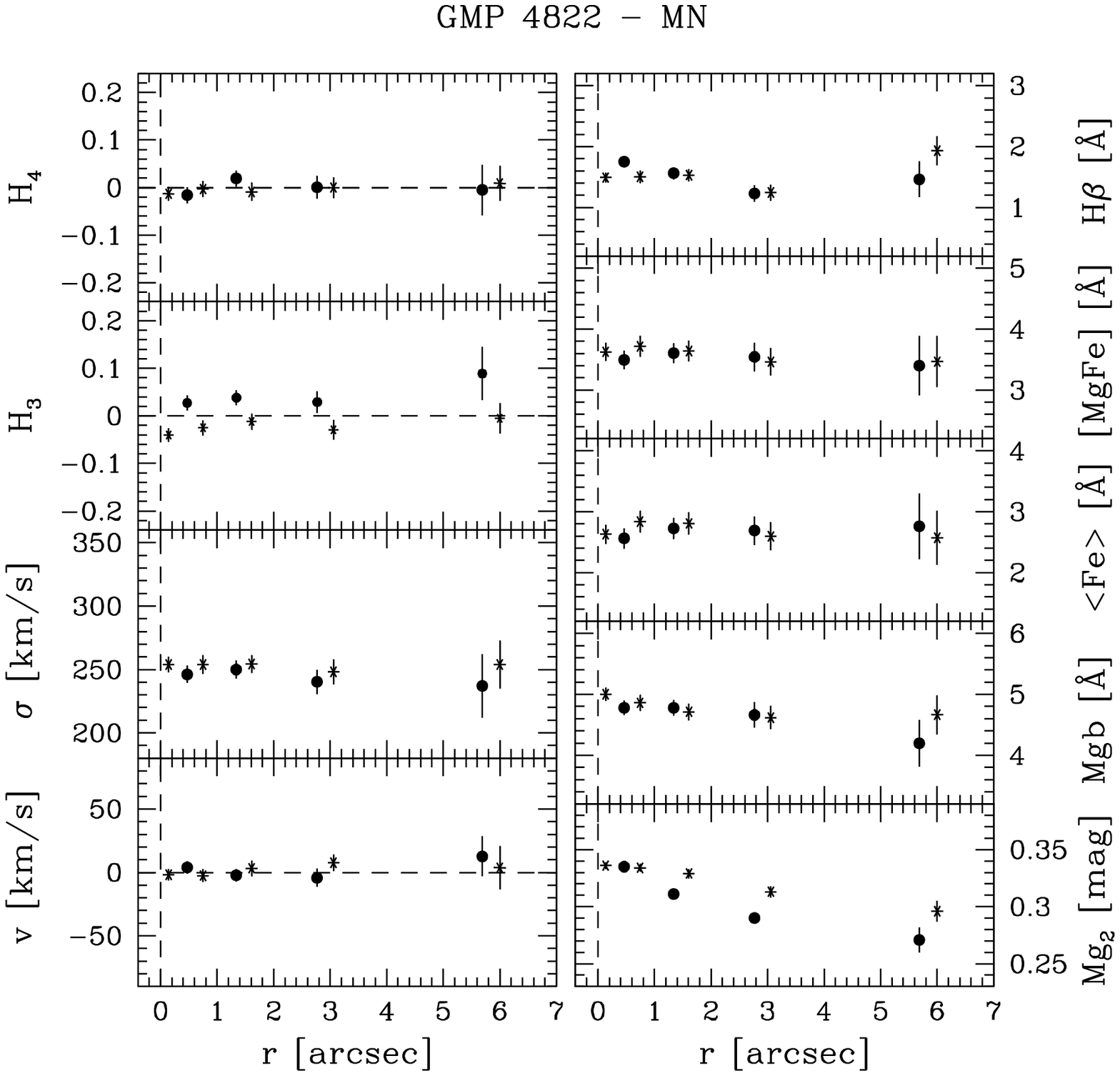}\\ 
\plottwo{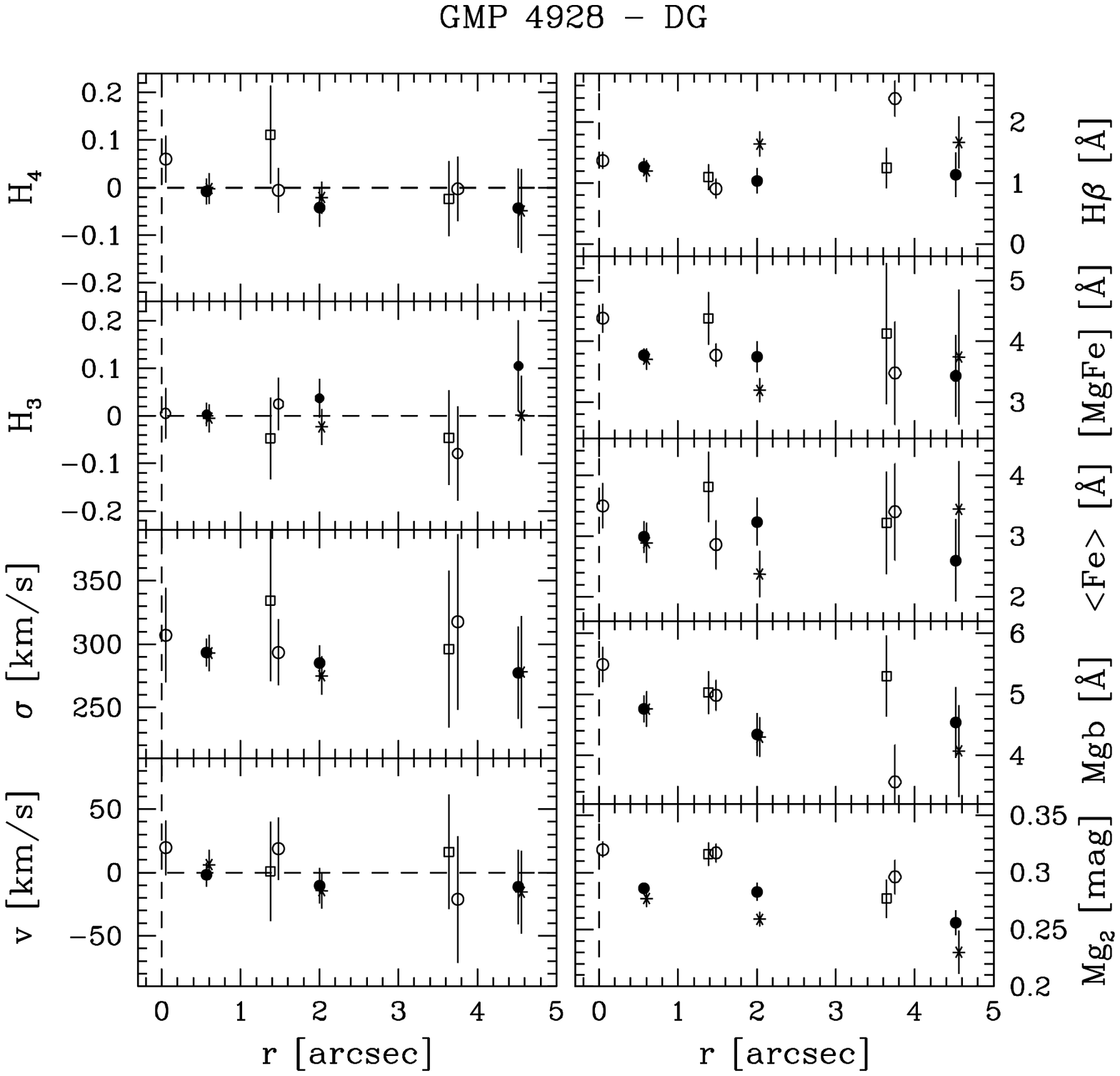}{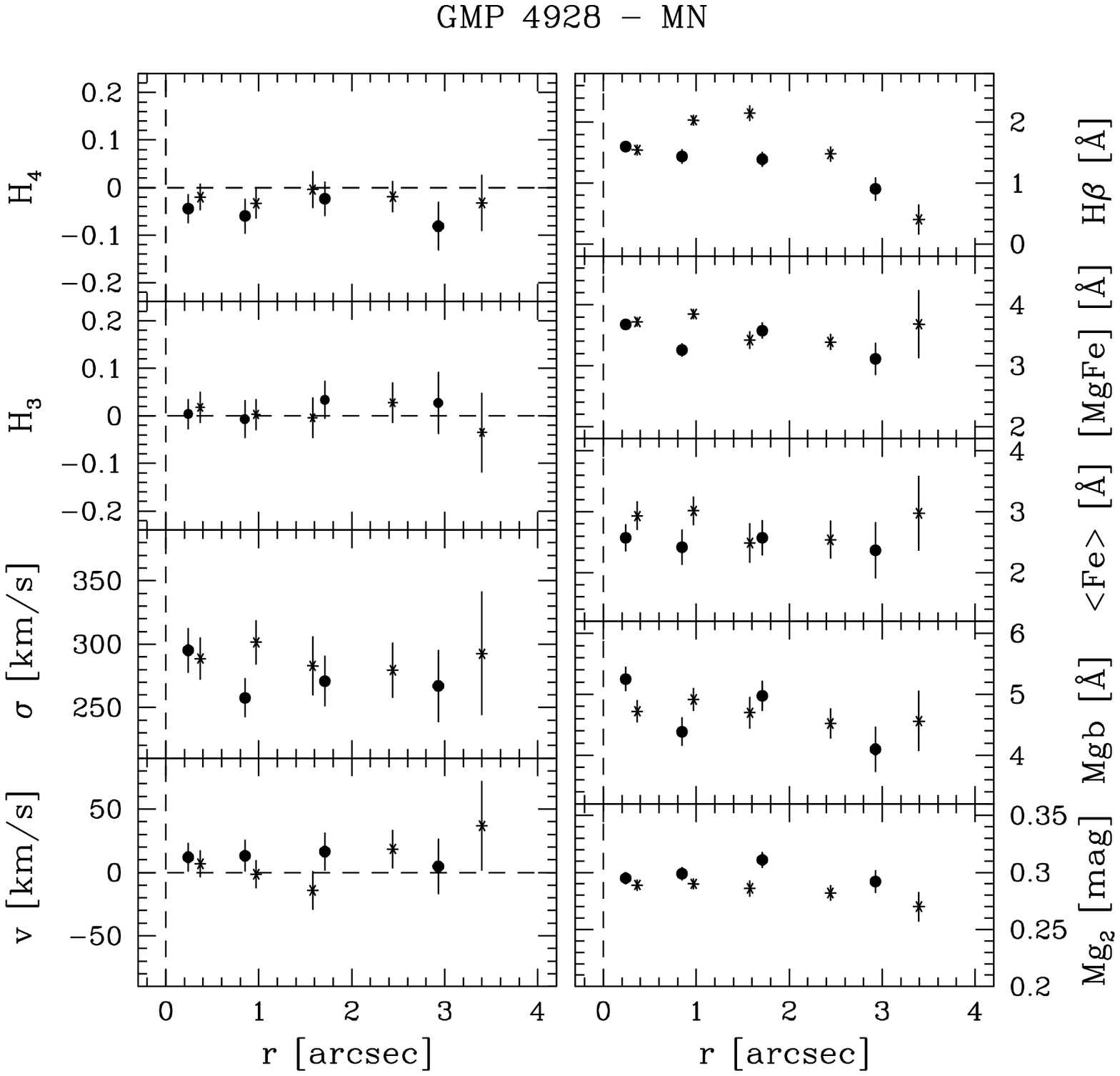}\\ 
\plottwo{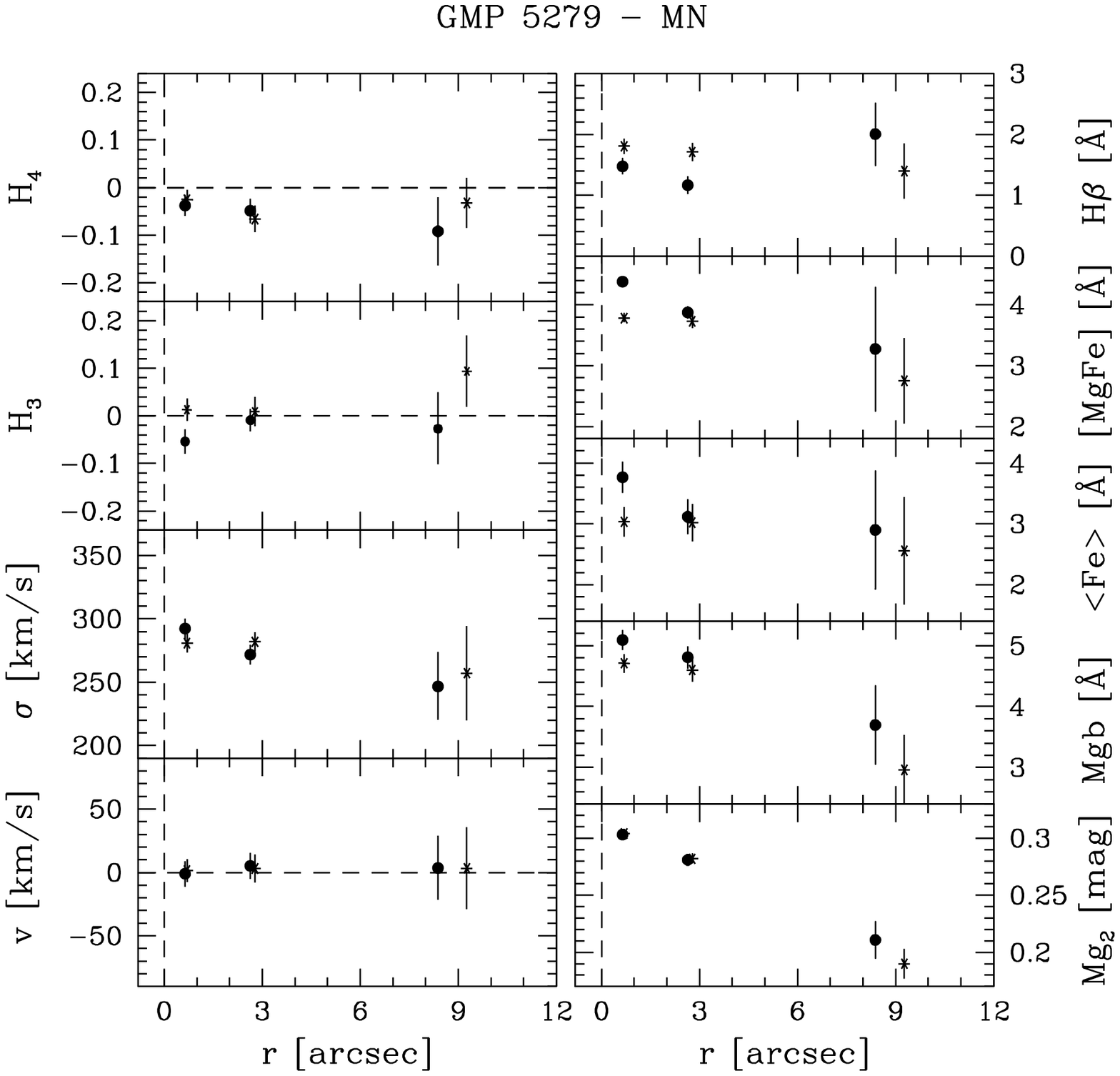}{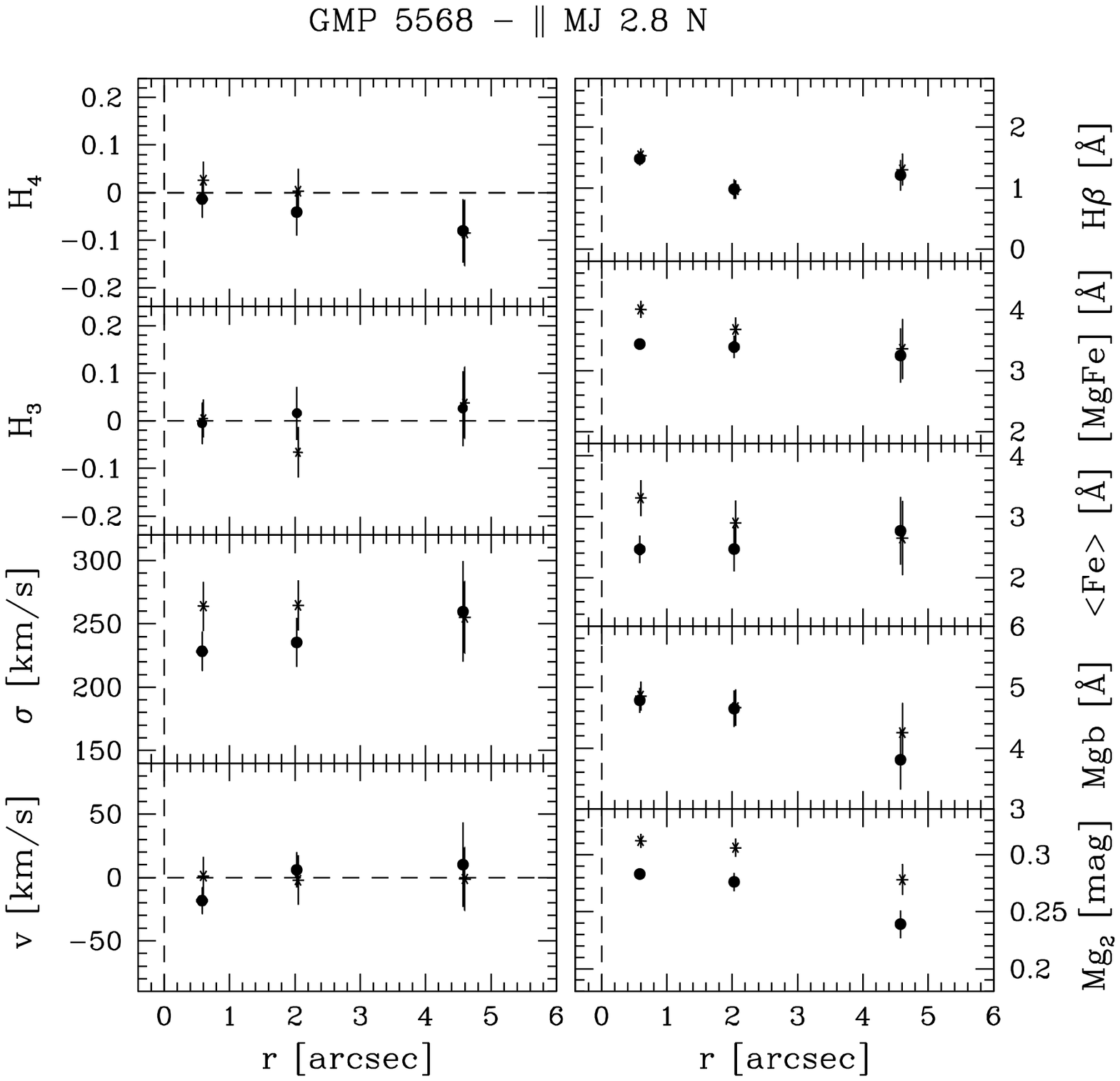} 
\caption{Continued.} 
%\label{fig:kinematics}   
\end{figure} 
 
%\clearpage 
 
\addtocounter{figure}{-1} 
\begin{figure} 
\epsscale{1.0} 
\plottwo{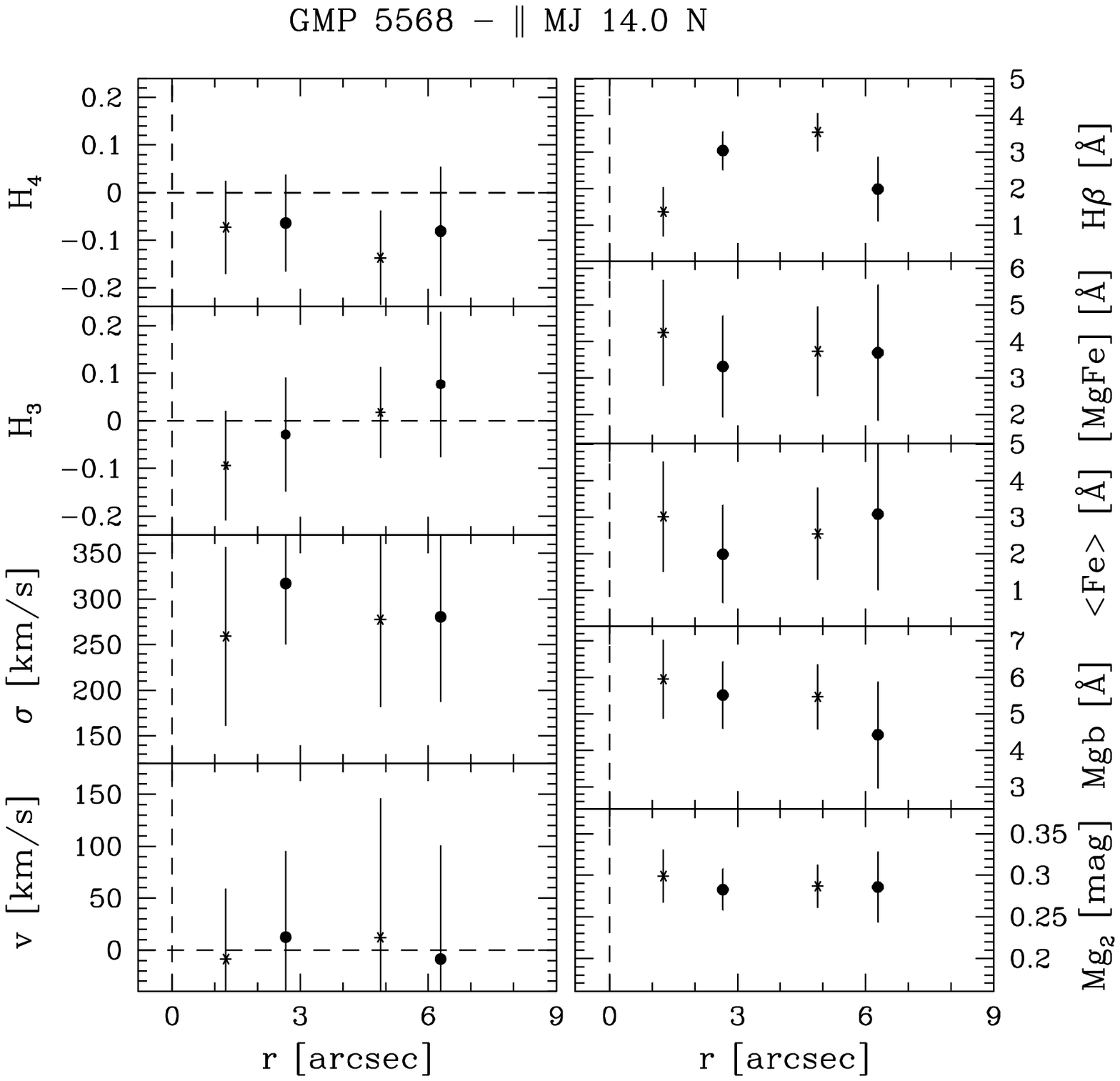}{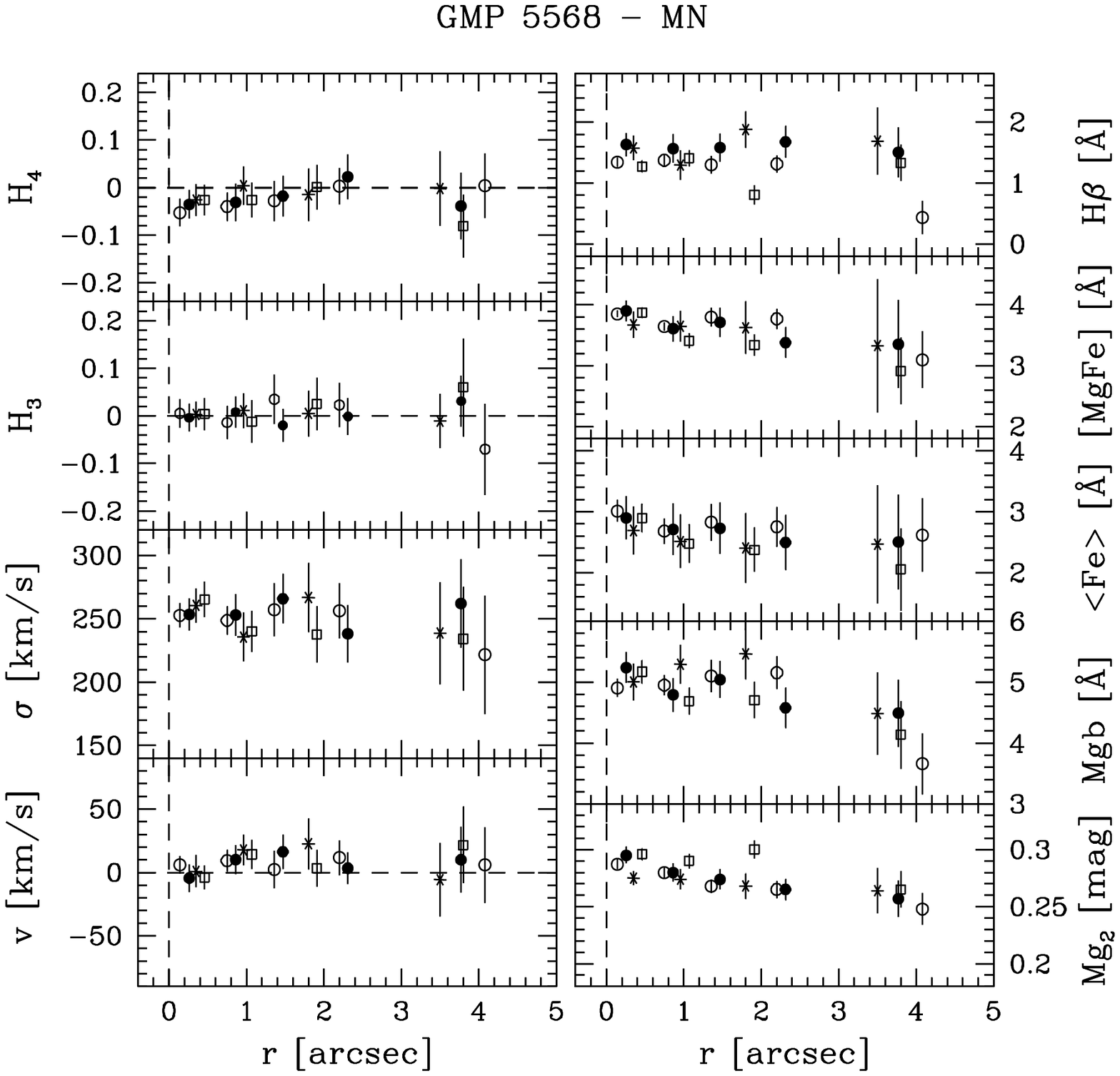}\\  
\plotone{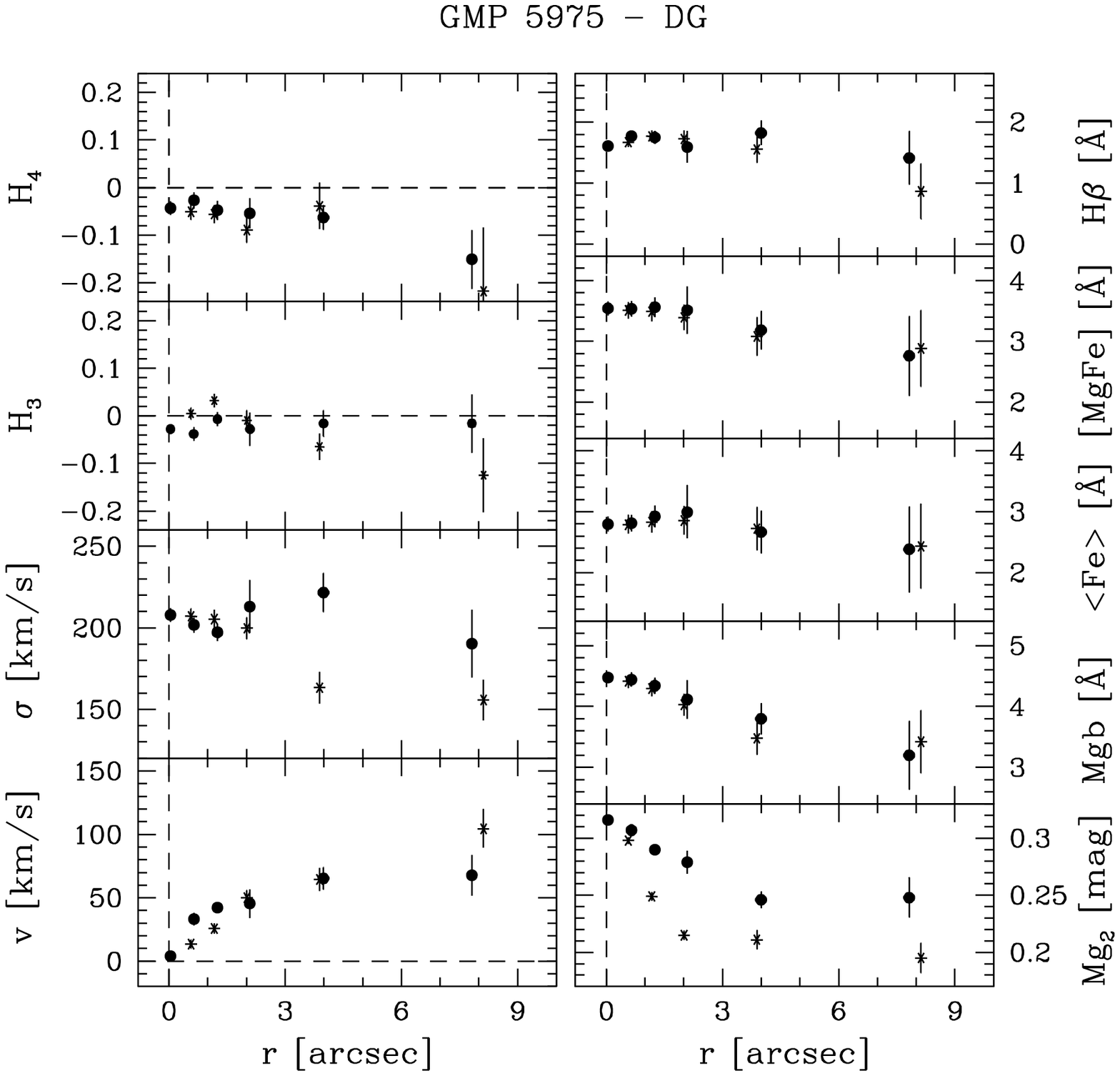}{ } 
\caption{Continued.} 
%\label{fig:kinematics}   
\end{figure} 
 
%\clearpage 
 
\begin{figure} 
\epsscale{.80} 
\plotone{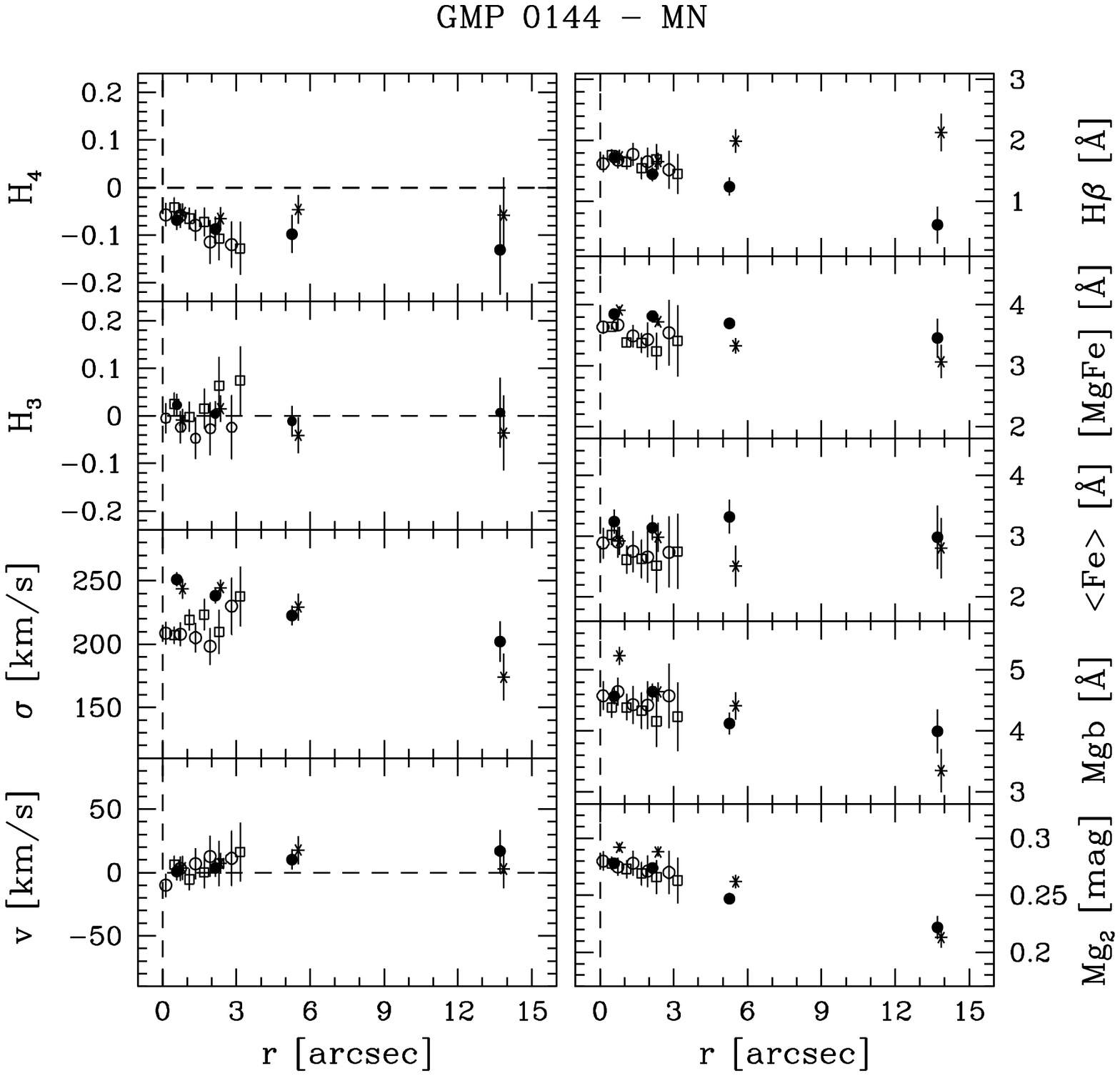} 
\caption{Kinematical parameters (left panels) and line strength 
  indices profiles (right panels) along the minor axis of GMP 0144 
  measured in run 1 (filled symbols) and in Paper II (open symbols). 
  Circles refer to data measured along the receding side of the 
  galaxy.} 
\label{fig:comparison}   
\end{figure} 
 
%\clearpage 
 
\begin{figure} 
\epsscale{.40} 
\includegraphics[scale=0.6,angle=270]{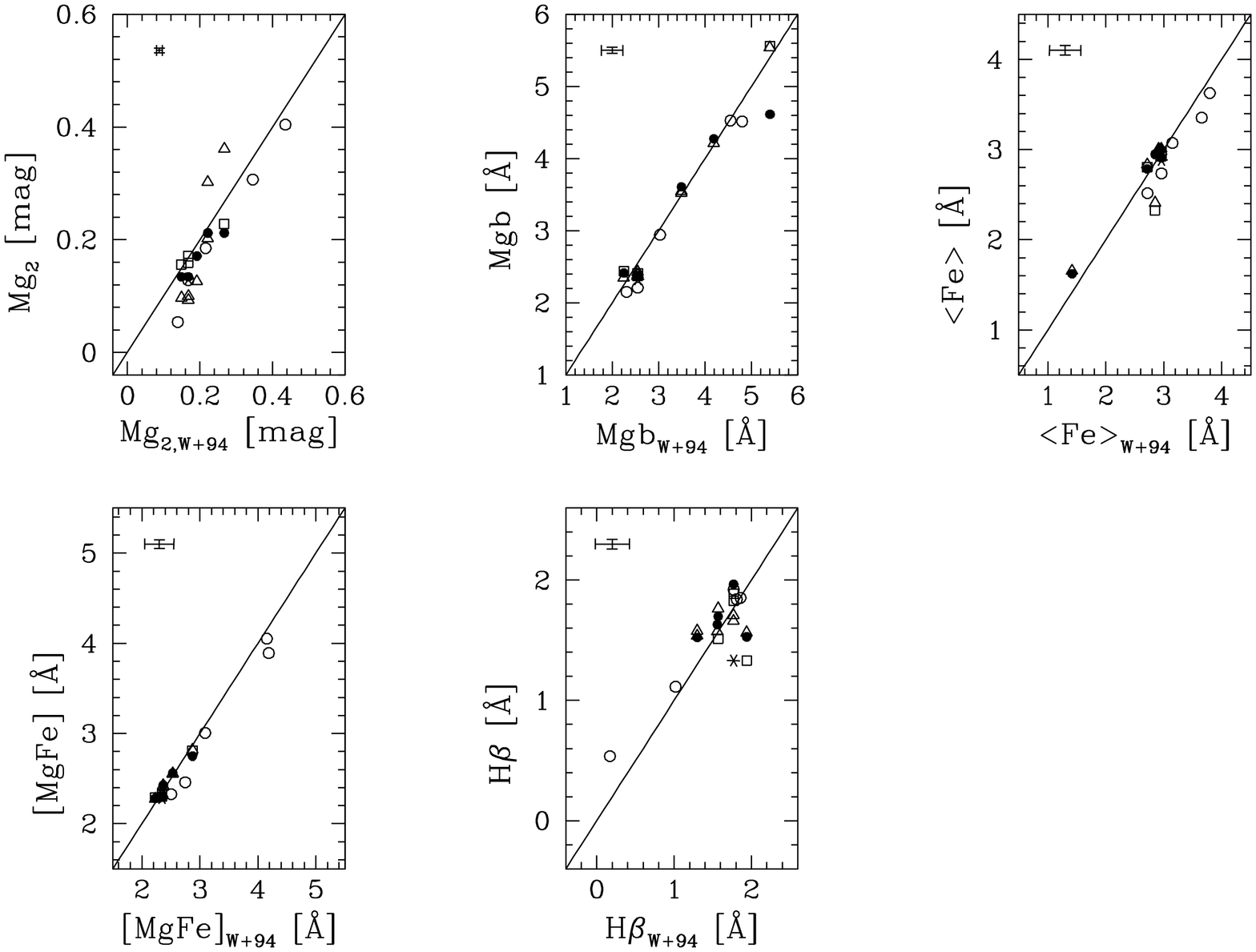} 
\caption{Comparison of the values for the line strength indices of \Mgd ,  
  \Mgb , \Mgd , \Fe , [MgFe], and \Hb\  measured for a sample 
  of stars in this work and in \citet[W$+$94]{worthey1994}.  The open 
  circles, filled circles, squares, asterisks, and triangles refer to 
  data obtained in Run 1, 2, 3, 4, and 5, respectively. In each panel 
  the error bars in the upper left corner indicate the mean errors and 
  the continous line shows the line of correspondence.} 
\label{fig:lick}   
\end{figure}

%\clearpage 
 
\begin{figure} 
\epsscale{.40} 
\includegraphics[scale=0.6,angle=270]{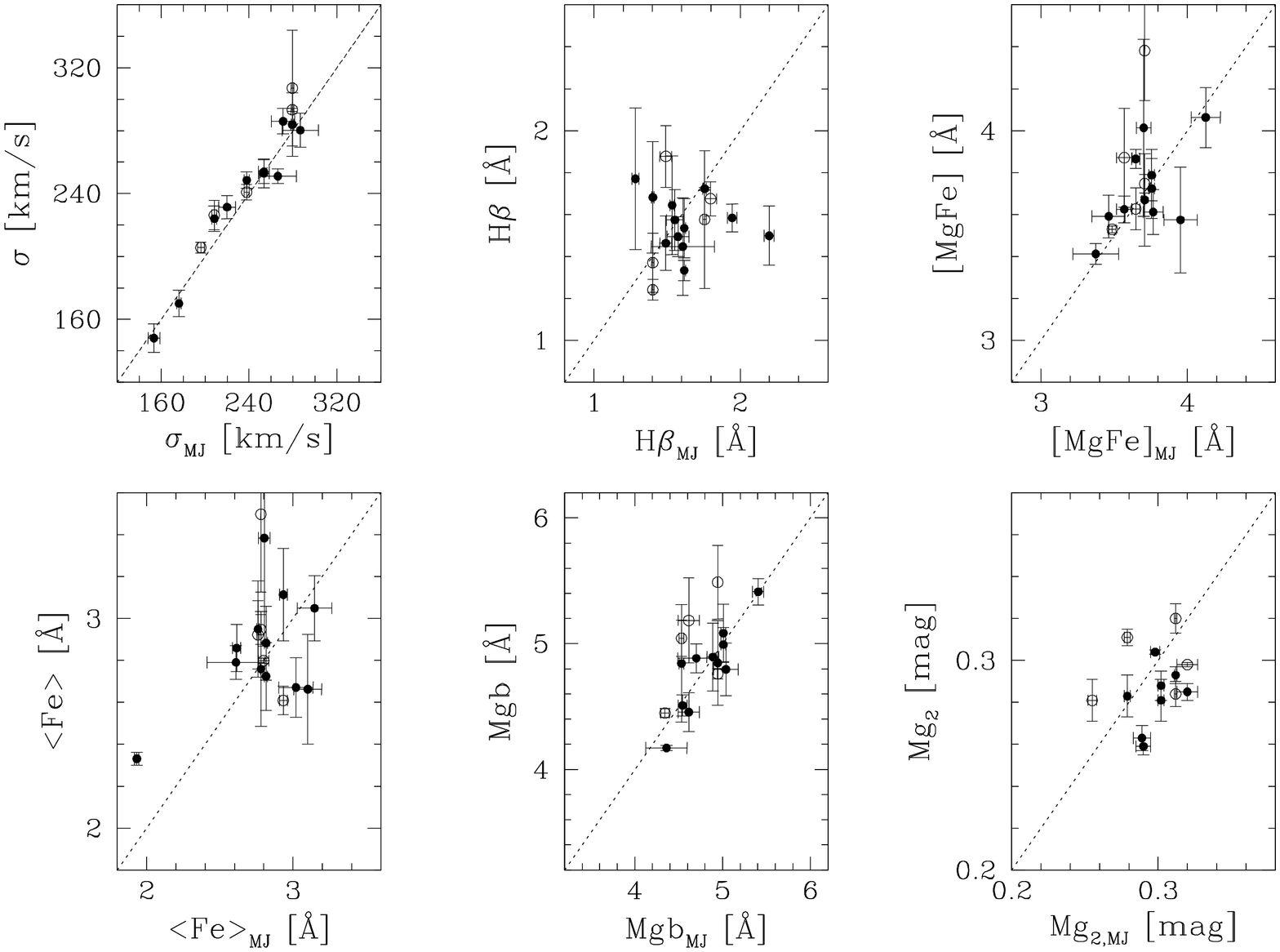} 
\caption{The central values of $\sigma$, \Hb , [MgFe], \Fe ,  
  \Mgb , and \Mgd\ measured within an aperture of $2$ arcsec along the 
  minor (filled circles) and diagonal axis (open circles) are compared 
  to those measured in Paper I along the major axis.} 
\label{fig:central} 
\end{figure} 
 
%\clearpage 
 
\begin{rotate} 
\begin{deluxetable}{rrrrrrcrrrrrrc} 
\tablecolumns{14} 
\tabletypesize{\tiny} 
\tablewidth{0pt} 
\tablecaption{Photometric parameters of the sample galaxies observed with HST/WFPC2\label{tab:photometry}} 
\tablehead{ 
\colhead{$a$} & \colhead{$\mu_R$} & \colhead{$e$} & \colhead{PA} &  
\colhead{$\Delta x_{\rm c}$} & \colhead{$\Delta y_{\rm c}$} & \colhead{Err.\tablenotemark{a}}& 
\colhead{$a_3/a$} & \colhead{$b_3/a$} &  
\colhead{$a_4/a$} & \colhead{$b_4/a$} & 
\colhead{$a_6/a$} & \colhead{$b_6/a$} & \colhead{Err.\tablenotemark{b}} \\ 
\colhead{[arcsec]} & \colhead{[mag arcsec$^{-2}$]} & \colhead{} & \colhead{[$^\circ$]} &  
\colhead{[arcsec]} & \colhead{[arcsec]} & \colhead{[arcsec]} & 
\colhead{$\times 100$} & \colhead{$\times 100$} & 
\colhead{$\times 100$} & \colhead{$\times 100$} & 
\colhead{$\times 100$} & \colhead{$\times 100$} & \colhead{}} 
\startdata 
\cutinhead{GMP 0756} 
$ 0.019\pm0.005$ & $14.013\pm0.008$ & $0.190\pm0.312$ & $ 79.8\pm58.1$ & $-0.001$ & $-0.005$ & 0.004 & $ -2.11$ & $ -7.78$ & $-1.54$ & $  0.87$ & $-0.81$ & $ -0.82$ & 0.87\\ 
$ 0.057\pm0.003$ & $14.182\pm0.015$ & $0.271\pm0.046$ & $ 85.0\pm 6.6$ & $ 0.000$ & $ 0.000$ & 0.002 & $ -2.59$ & $ -4.66$ & $ 0.11$ & $  1.10$ & $ 0.10$ & $ -0.76$ & 0.51\\ 
$ 0.073\pm0.001$ & $14.306\pm0.015$ & $0.256\pm0.018$ & $ 84.9\pm 2.7$ & $-0.001$ & $ 0.000$ & 0.001 & $ -1.85$ & $ -2.42$ & $ 0.15$ & $  0.20$ & $ 0.19$ & $ -0.73$ & 0.40\\ 
$ 0.088\pm0.001$ & $14.418\pm0.012$ & $0.240\pm0.011$ & $ 84.9\pm 1.7$ & $-0.002$ & $-0.000$ & 0.001 & $ -1.42$ & $ -1.31$ & $ 0.23$ & $ -0.46$ & $ 0.17$ & $ -0.56$ & 0.37\\ 
$ 0.103\pm0.001$ & $14.504\pm0.015$ & $0.250\pm0.010$ & $ 84.8\pm 1.6$ & $-0.003$ & $-0.001$ & 0.001 & $ -1.54$ & $ -0.78$ & $ 1.08$ & $ -0.93$ & $ 0.59$ & $ -0.34$ & 0.37\\ 
$ 0.124\pm0.001$ & $14.608\pm0.012$ & $0.297\pm0.009$ & $ 86.7\pm 1.2$ & $-0.003$ & $-0.002$ & 0.001 & $ -1.90$ & $ -0.84$ & $ 0.56$ & $ -0.28$ & $-0.54$ & $  0.32$ & 0.39\\ 
$ 0.143\pm0.001$ & $14.729\pm0.013$ & $0.304\pm0.008$ & $ 88.1\pm 1.1$ & $-0.003$ & $-0.003$ & 0.001 & $ -1.87$ & $ -0.94$ & $-0.33$ & $  0.24$ & $ 0.05$ & $  0.78$ & 0.36\\ 
$ 0.162\pm0.001$ & $14.839\pm0.014$ & $0.305\pm0.006$ & $ 88.2\pm 0.8$ & $-0.003$ & $-0.003$ & 0.001 & $ -1.26$ & $ -0.59$ & $ 0.68$ & $  0.38$ & $-0.10$ & $  0.46$ & 0.28\\
\enddata 
\tablenotetext{a}{Error on the center coordinates derived from the residual rms 
of the ellipse fit to the isophotes: Err$={\rm 
    rms_{fit}}/\sqrt{N}$ with $N\leq128$ the number of fitted points 
  of the isophote.}   
\tablenotetext{b}{Error of Fourier coefficients 
  defined as 
  Err$=\sqrt{\frac{\sum_{i=10}^{N/2}\,\left(a_i^2+b_i^2\right)}{N/2-10}}\times\frac{100}{a}$.} 
\end{deluxetable} 
\end{rotate}

\clearpage		   	 			                 
 
\begin{deluxetable}{crrrrrrrrr}	 			                 
\tablecolumns{10}  
\tabletypesize{\tiny}  
\tablewidth{0pt} 
\tablecaption{Second, eighth, tenth and twelfth order cosine Fourier 
   coefficients of the isophotal shape decomposition of 
   GMP1176\label{tab:phot1176}}  
\tablehead{  
\colhead{$a$} & \colhead{$a_2/a$} & \colhead{$a_{8}/a$} & 
\colhead{$a_{10}/a$} & \colhead{$a_{12}/a$} &  
\colhead{$a$} & \colhead{$a_2/a$} & \colhead{$a_{8}/a$} & 
\colhead{$a_{10}/a$} & \colhead{$a_{12}/a$} \\ 
\colhead{[arcsec]} & \colhead{$\times 100$} & \colhead{$\times 100$} &  
\colhead{$\times 100$} & \colhead{$\times 100$} & 
\colhead{[arcsec]} & \colhead{$\times 100$} & \colhead{$\times 100$} &  
\colhead{$\times 100$} & \colhead{$\times 100$} } 
\startdata 
$0.015$ & $-0.05$ & $-0.34$ & $-0.17$ & $-0.02$ & $ 1.489$ & $ 2.42$ & $ 2.92$ & $ 1.63$ & $ 1.09$\\       
$0.044$ & $-0.17$ & $ 0.43$ & $ 0.34$ & $-0.06$ & $ 1.604$ & $ 2.54$ & $ 2.85$ & $ 1.58$ & $ 0.94$\\         
$0.059$ & $-0.13$ & $ 0.66$ & $ 0.17$ & $-0.17$ & $ 1.741$ & $ 2.78$ & $ 2.43$ & $ 1.21$ & $ 0.71$\\         
$0.073$ & $-0.05$ & $ 0.53$ & $ 0.05$ & $-0.13$ & $ 1.874$ & $ 2.90$ & $ 2.13$ & $ 0.99$ & $ 0.85$\\         
$0.086$ & $ 0.05$ & $ 0.48$ & $ 0.09$ & $-0.16$ & $ 2.005$ & $ 2.90$ & $ 2.08$ & $ 1.05$ & $ 0.90$\\         
$0.097$ & $ 0.12$ & $ 0.54$ & $ 0.13$ & $-0.16$ & $ 2.254$ & $ 3.01$ & $ 2.27$ & $ 1.40$ & $ 1.07$\\         
$0.111$ & $ 0.23$ & $-0.12$ & $-0.19$ & $-0.10$ & $ 2.526$ & $ 3.12$ & $ 2.89$ & $ 1.95$ & $ 1.27$\\         
$0.126$ & $ 0.16$ & $ 0.19$ & $-0.07$ & $-0.15$ & $ 2.803$ & $ 3.18$ & $ 3.19$ & $ 2.05$ & $ 1.13$\\
\enddata 
\end{deluxetable}

\begin{table} 
\begin{scriptsize} 
\caption{Log of spectroscopic observations \label{tab:log}}    
\begin{tabular}{lcrccrccc}   
\tableline 
\tableline  
\multicolumn{1}{c}{Object} &   
\multicolumn{1}{c}{Run} & \multicolumn{1}{c}{P.A.} &   
\multicolumn{1}{c}{Position} & \multicolumn{1}{c}{Offset} &   
\multicolumn{1}{c}{Single Exp. Time}& \multicolumn{1}{c}{Total Exp. Time} & 
\multicolumn{1}{c}{FWHM} & \multicolumn{1}{c}{Q}\\  
\multicolumn{1}{c}{} &   
\multicolumn{1}{c}{} & \multicolumn{1}{c}{[$^\circ$]} &   
\multicolumn{1}{c}{} & \multicolumn{1}{c}{[arcsec]} &   
\multicolumn{1}{c}{[s]} & \multicolumn{1}{c}{[h]} & 
\multicolumn{1}{c}{}& \multicolumn{1}{c}{[arcsec]} \\  
\multicolumn{1}{c}{(1)} & \multicolumn{1}{c}{(2)} &  
\multicolumn{1}{c}{(3)} & \multicolumn{1}{c}{(4)} &  
\multicolumn{1}{c}{(5)} & \multicolumn{1}{c}{(6)} &  
\multicolumn{1}{c}{(7)} & \multicolumn{1}{c}{(7)} &  
\multicolumn{1}{c}{(9)} \\  
\tableline   
\object{GMP 0144} & 1 & 46  & DG             & \nodata& $1\times2700$      & 0.8 &    2.2 & 2\\ 
                  & 1 & 1   & MN             & \nodata& $1\times2700$      & 0.8 &    1.8 & 1\\          
\object{GMP 2417} & 3 & 55  & $\parallel$ MJ & 3.8 S  & $6\times3600$      & 6.0 &    1.7 & 3\\ 
                  & 3 & 145 & MN             & \nodata& $3\times3600$      & 3.0 &    2.3 & 3\\ 
\object{GMP 2440} & 1 & 75  & DG             & \nodata& $1\times2700$      & 0.8 &    3.2 & 2\\
\end{tabular}    
\end{scriptsize} 
\tablecomments{   
Col. 1: GMP No. from Godwin, Metcalfe \& Peach (1983). 
Col. 2: Observing run.  
Col. 3: Slit position angle measured North through East.  
Col. 4: Slit position. MJ = major axis; $\parallel$ MJ =   
        parallel to major axis; MN = minor axis; DG = diagonal axis.  
Col. 5: Northward/soutward offset of the slit with respect   
        to galaxy center.   
Col. 6: Number and exposure time of the single exposures.  
Col. 7: Total exposure time.  
Col. 8: FWHM of the seeing and instrumental PSF of the resulting spectrum.
Col. 9: Estimated quality of the resulting spectrum.  
         1: very good; 2: good; 3: medium (see Fig.  
         \ref{fig:quality}).} 
\end{table}   
 
\clearpage 
 
\begin{rotate} 
\begin{table} 
\begin{scriptsize} 
\caption{Instrumental set-up of spectroscopic observations \label{tab:setup}} 
\begin{tabular}{l ll ll l} 
\tableline 
\tableline 
Parameter                     & Run 1                         & Run 2                       & Run 3                       & Run 4                           & Run 5                        \\ 
\tableline 
Telescope                     & HET                           & 2.4-m Hiltner               & 2.4-m Hiltner               & HET                             & 2.4-m Hiltner               \\ 
Spectrograph                  & LRS                           & Modular                     & Modular                     & LRS                             & Modular                     \\ 
Grating                       & Grism G2 600 $\rm gr\,mm^{-1}$& 1200  $\rm gr\,mm^{-1}$     & 1200  $\rm gr\,mm^{-1}$     & Echelle E2 316 $\rm gr\,mm^{-1}$& 1200  $\rm gr\,mm^{-1}$     \\ 
CCD                           & Ford Aerospace                & ``Echelle'' SITe            & ``Echelle'' SITe            & Ford Aerospace                  & ``Echelle'' SITe            \\ 
Pixel number                  & $3072\times3072$              & $2048\times2048$            & $2048\times2048$            & $3072\times3072$                & $2048\times2048$            \\ 
Pixel size                    & $15\times15$ $\rm \mu\;m^2$   & $24\times24$ $\rm \mu\;m^2$ & $24\times24$ $\rm \mu\;m^2$ & $15\times15$ $\rm \mu\;m^2$     & $24\times24$ $\rm \mu\;m^2$ \\ 
Binning                       & $2\times2$                    & $1\times1$                  & $1\times1$                  & $2\times2$                      & $1\times1$                  \\ 
Gain                          & 1.8 $e^{-}{\rm ADU}^{-1}$     & 2.7 $e^{-}{\rm ADU}^{-1}$   & 2.7 $e^{-}{\rm ADU}^{-1}$   & 1.8 $e^{-}{\rm ADU}^{-1}$       & 2.7 $e^{-}{\rm ADU}^{-1}$   \\ 
RON                           & 5.2 $e^{-}$                   & 7.9 $e^{-}$                 & 7.9 $e^{-}$                 & 5.2 $e^{-}$                     & 7.9 $e^{-}$                 \\ 
Scale                         & $\rm 0\farcs47\;pixel^{-1}$   & $\rm 0\farcs606\;pixel^{-1}$& $\rm 0\farcs606\;pixel^{-1}$& $\rm 0\farcs47\;pixel^{-1}$     & $\rm 0\farcs606\;pixel^{-1}$\\ 
Dispersion                    & $\rm 1.99\;\AA\;pixel^{-1}$   & $\rm 0.91\;\AA\;pixel^{-1}$ & $\rm 1.01\;\AA\;pixel^{-1}$ & $\rm 0.72\;\AA\;pixel^{-1}$     & $\rm 1.01\;\AA\;pixel^{-1}$ \\ 
Slit width                    & $1\farcs0$                    & $1\farcs9$                  & $1\farcs9$                  & $1\farcs0$                      & $1\farcs9$                  \\ 
Wavelength range\tablenotemark{a}  & 4284 -- 7400 \AA         & 4702 -- 6401 \AA            & 4644 -- 6455 \AA            & 4870 -- 5956 \AA                & 4544 -- 6363 \AA            \\ 
Instrumental FWHM             & $4.7$ \AA                     & $2.5$ \AA                   & $2.3$ \AA                   & $2.1$ \AA                       & $2.3$ \AA                   \\ 
Instrumental $\sigma$\tablenotemark{b} & $114$ \kms           & $61$ \kms                   & $53$ \kms                   & $51$ \kms                       & $53$ \kms                   \\ 
\tableline 
\end{tabular} 
\end{scriptsize} 
\tablenotetext{a}{Measured on reduced spectra.} 
\tablenotetext{b}{Derived at the 5170 \AA\ (corresponding to \MgI\ triplet).} 
\end{table} 
\end{rotate} 
 
\clearpage 
			   	 			                 
\begin{deluxetable}{rrrrrrrr}	 			                 
\tablecolumns{8}	   	 			                 
\tabletypesize{\scriptsize} 
\tablewidth{0pt} 
\tablecaption{Stellar kinematics of the sample galaxies \label{tab:kinematics}} 
\tablehead{ 
\colhead{$r$} & \colhead{$V$} & \colhead{$\sigma$} &  
\colhead{$H_3$} & \colhead{$H_4$} &  
\colhead{PA} & \colhead{Offset} & \colhead{Run}\\ 
\colhead{[arcsec]} & \colhead{[\kms]} & \colhead{[\kms]} &  
\colhead{} & \colhead{} &  
\colhead{[$^\circ$]} & \colhead{[arcsec]} & \colhead{}} 
\startdata 
\cutinhead{GMP 0144} 
$ -3.83$ & $  -7.9\pm 11.8$ & $248.8\pm11.3$ & $-0.044\pm0.039$ & $-0.063\pm0.034$ &  46 & $ 0.0$ & 1\\ 
$ -0.88$ & $  -9.9\pm  7.6$ & $244.0\pm 6.8$ & $ 0.003\pm0.021$ & $-0.034\pm0.021$ &  46 & $ 0.0$ & 1\\ 
$  0.83$ & $   4.8\pm  7.1$ & $237.0\pm 7.7$ & $-0.023\pm0.026$ & $-0.045\pm0.025$ &  46 & $ 0.0$ & 1\\ 
$  3.90$ & $  12.9\pm 16.7$ & $225.3\pm17.1$ & $-0.008\pm0.052$ & $-0.082\pm0.051$ &  46 & $ 0.0$ & 1\\ 
$-13.86$ & $  -2.7\pm 15.1$ & $173.9\pm18.5$ & $ 0.036\pm0.079$ & $-0.058\pm0.080$ &   1 & $ 0.0$ & 1\\  
$ -5.52$ & $ -17.6\pm 11.1$ & $229.3\pm10.7$ & $ 0.041\pm0.038$ & $-0.046\pm0.030$ &   1 & $ 0.0$ & 1\\ 
$ -2.37$ & $  -7.1\pm  8.3$ & $244.5\pm 6.5$ & $-0.015\pm0.028$ & $-0.065\pm0.024$ &   1 & $ 0.0$ & 1\\ 
$ -0.79$ & $  -3.5\pm  9.1$ & $243.6\pm 7.4$ & $ 0.009\pm0.023$ & $-0.053\pm0.019$ &   1 & $ 0.0$ & 1\\ 
$  0.57$ & $   0.5\pm  6.9$ & $250.9\pm 5.2$ & $ 0.023\pm0.023$ & $-0.068\pm0.021$ &   1 & $ 0.0$ & 1\\ 
$  2.14$ & $   3.4\pm  6.7$ & $238.2\pm 6.2$ & $ 0.004\pm0.027$ & $-0.087\pm0.026$ &   1 & $ 0.0$ & 1\\ 
$  5.27$ & $  10.2\pm  8.0$ & $222.6\pm 7.7$ & $-0.011\pm0.032$ & $-0.098\pm0.039$ &   1 & $ 0.0$ & 1\\ 
$ 13.71$ & $  16.9\pm 16.5$ & $202.0\pm16.0$ & $ 0.007\pm0.073$ & $-0.131\pm0.094$ &   1 & $ 0.0$ & 1\\ 
\cutinhead{GMP 2417} 
$ -6.92$ & $ -90.9\pm 43.7$ & $172.7\pm42.6$ & $ 0.014\pm0.105$ & $-0.096\pm0.134$ &  55 & $-3.8$ & 3\\ 
$ -2.56$ & $ -87.2\pm 34.1$ & $203.5\pm39.9$ & $ 0.001\pm0.108$ & $-0.059\pm0.080$ &  55 & $-3.8$ & 3\\ 
$  1.43$ & $  77.7\pm 37.1$ & $217.5\pm48.6$ & $ 0.068\pm0.101$ & $ 0.027\pm0.088$ &  55 & $-3.8$ & 3\\ 
$  6.23$ & $ 100.5\pm 49.5$ & $175.9\pm62.1$ & $ 0.044\pm0.152$ & $-0.011\pm0.141$ &  55 & $-3.8$ & 3\\ 
$ -3.29$ & $ -32.1\pm 25.6$ & $175.1\pm33.0$ & $ 0.052\pm0.079$ & $-0.019\pm0.110$ & 145 & $ 0.0$ & 3\\  
$ -1.84$ & $  10.2\pm 14.3$ & $233.4\pm15.9$ & $-0.014\pm0.039$ & $-0.026\pm0.033$ & 145 & $ 0.0$ & 3\\  
$ -1.00$ & $  -4.4\pm 11.5$ & $221.0\pm11.2$ & $-0.004\pm0.036$ & $-0.089\pm0.031$ & 145 & $ 0.0$ & 3\\  
$ -0.40$ & $  18.4\pm  9.4$ & $237.2\pm 9.9$ & $ 0.024\pm0.029$ & $-0.049\pm0.024$ & 145 & $ 0.0$ & 3\\  
$  0.21$ & $ -13.8\pm  9.7$ & $235.8\pm11.3$ & $ 0.005\pm0.030$ & $-0.034\pm0.026$ & 145 & $ 0.0$ & 3\\  
$  0.81$ & $   2.5\pm 11.2$ & $230.0\pm11.4$ & $ 0.014\pm0.034$ & $-0.059\pm0.029$ & 145 & $ 0.0$ & 3\\  
$  1.66$ & $   6.5\pm 12.4$ & $204.2\pm15.0$ & $-0.051\pm0.044$ & $-0.088\pm0.041$ & 145 & $ 0.0$ & 3\\  
$  3.66$ & $  13.0\pm 20.6$ & $167.7\pm21.8$ & $-0.024\pm0.109$ & $-0.115\pm0.096$ & 145 & $ 0.0$ & 3\\  
\cutinhead{GMP 2440} 
$ -3.00$ & $-123.0\pm 11.0$ & $158.2\pm12.9$ & $ 0.072\pm0.096$ & $-0.020\pm0.088$ &  75 & $ 0.0$ & 1\\  
$ -0.76$ & $ -64.3\pm  7.9$ & $219.3\pm 8.7$ & $ 0.007\pm0.038$ & $-0.079\pm0.056$ &  75 & $ 0.0$ & 1\\  
$  0.35$ & $  -0.2\pm  7.7$ & $232.6\pm 8.3$ & $-0.017\pm0.031$ & $-0.033\pm0.037$ &  75 & $ 0.0$ & 1\\  
$  1.45$ & $ 105.0\pm 13.1$ & $193.6\pm18.3$ & $-0.083\pm0.056$ & $ 0.080\pm0.053$ &  75 & $ 0.0$ & 1\\  
$  4.27$ & $ 132.7\pm 30.9$ & $145.9\pm18.2$ & $-0.116\pm0.121$ & $-0.056\pm0.126$ &  75 & $ 0.0$ & 1\\  
$ -4.36$ & $ -17.0\pm 24.8$ & $183.1\pm23.4$ & $ 0.072\pm0.120$ & $ 0.016\pm0.096$ &  18 & $ 0.0$ & 1\\  
$ -1.15$ & $   9.0\pm  8.6$ & $222.2\pm10.0$ & $ 0.029\pm0.038$ & $-0.031\pm0.042$ &  18 & $ 0.0$ & 1\\  
$  0.88$ & $   7.0\pm  7.1$ & $224.0\pm 8.2$ & $-0.011\pm0.032$ & $-0.050\pm0.042$ &  18 & $ 0.0$ & 1\\  
$  4.06$ & $   1.1\pm 28.4$ & $180.5\pm19.3$ & $-0.044\pm0.114$ & $-0.072\pm0.118$ &  18 & $ 0.0$ & 1\\  
\cutinhead{GMP 2921}										         
$-11.08$ & $ -20.5\pm 52.3$ & $412.8\pm81.7$ & $-0.015\pm0.095$ & $ 0.034\pm0.094$ & 171 & $ 0.0$ & 5\\ 
$ -5.90$ & $  12.5\pm 13.1$ & $356.2\pm17.0$ & $ 0.007\pm0.024$ & $ 0.038\pm0.029$ & 171 & $ 0.0$ & 5\\ 
$ -2.60$ & $   1.3\pm 10.2$ & $391.5\pm14.3$ & $ 0.004\pm0.018$ & $ 0.030\pm0.022$ & 171 & $ 0.0$ & 5\\ 
$ -1.16$ & $  11.7\pm  7.5$ & $399.4\pm10.4$ & $ 0.017\pm0.013$ & $ 0.026\pm0.016$ & 171 & $ 0.0$ & 5\\ 
$  0.03$ & $  10.4\pm  6.5$ & $390.3\pm 8.3$ & $-0.001\pm0.012$ & $-0.003\pm0.014$ & 171 & $ 0.0$ & 5\\ 
$  1.21$ & $  -5.4\pm  9.1$ & $383.9\pm 8.0$ & $ 0.023\pm0.015$ & $-0.022\pm0.012$ & 171 & $ 0.0$ & 5\\ 
$  2.65$ & $  -8.6\pm  8.7$ & $384.7\pm11.4$ & $ 0.008\pm0.016$ & $ 0.003\pm0.019$ & 171 & $ 0.0$ & 5\\ 
$  5.76$ & $  -4.6\pm 15.3$ & $414.0\pm18.2$ & $-0.006\pm0.025$ & $ 0.018\pm0.021$ & 171 & $ 0.0$ & 5\\ 
$ 10.76$ & $   3.3\pm 45.8$ & $388.1\pm63.8$ & $ 0.019\pm0.072$ & $ 0.001\pm0.054$ & 171 & $ 0.0$ & 5\\ 
\cutinhead{GMP 3329}										         
$ -9.50$ & $ -17.3\pm 35.9$ & $269.3\pm58.7$ & $ 0.008\pm0.081$ & $ 0.049\pm0.073$ & 135 & $ 0.0$ & 5\\ 
$ -4.86$ & $   6.8\pm 12.6$ & $246.5\pm16.0$ & $ 0.010\pm0.029$ & $-0.009\pm0.034$ & 135 & $ 0.0$ & 5\\ 
$ -2.55$ & $  -0.9\pm  8.7$ & $272.3\pm11.5$ & $ 0.000\pm0.020$ & $-0.007\pm0.022$ & 135 & $ 0.0$ & 5\\ 
$ -1.08$ & $   1.8\pm  8.3$ & $268.6\pm 9.8$ & $-0.004\pm0.018$ & $-0.025\pm0.025$ & 135 & $ 0.0$ & 5\\ 
$  0.11$ & $  -7.1\pm  7.9$ & $280.3\pm10.9$ & $ 0.002\pm0.019$ & $-0.028\pm0.028$ & 135 & $ 0.0$ & 5\\ 
$  1.31$ & $   2.9\pm  7.8$ & $269.4\pm10.9$ & $-0.020\pm0.020$ & $-0.025\pm0.027$ & 135 & $ 0.0$ & 5\\ 
$  2.78$ & $   5.2\pm  8.4$ & $243.1\pm10.5$ & $ 0.003\pm0.020$ & $-0.029\pm0.025$ & 135 & $ 0.0$ & 5\\ 
$  5.08$ & $   3.1\pm  9.6$ & $219.9\pm15.0$ & $-0.012\pm0.036$ & $-0.050\pm0.032$ & 135 & $ 0.0$ & 5\\ 
$  9.56$ & $   5.8\pm 33.2$ & $274.5\pm31.5$ & $-0.056\pm0.058$ & $-0.087\pm0.053$ & 135 & $ 0.0$ & 5\\ 
\cutinhead{GMP 3414}										         
$ -6.88$ & $ -88.3\pm 37.3$ & $ 92.6\pm56.9$ & $ 0.079\pm0.214$ & $-0.119\pm0.206$ & 178 & $ 4.6$ & 3\\ 
$ -2.47$ & $ -49.2\pm 31.1$ & $158.7\pm34.3$ & $ 0.004\pm0.094$ & $-0.158\pm0.088$ & 178 & $ 4.6$ & 3\\ 
$  1.56$ & $  33.4\pm 34.8$ & $159.8\pm34.9$ & $ 0.001\pm0.143$ & $-0.083\pm0.093$ & 178 & $ 4.6$ & 3\\ 
$  6.01$ & $ 104.1\pm 49.7$ & $130.8\pm74.2$ & $-0.183\pm0.259$ & $ 0.035\pm0.102$ & 178 & $ 4.6$ & 3\\ 
$ -2.99$ & $  11.8\pm 22.1$ & $158.0\pm22.8$ & $ 0.018\pm0.111$ & $-0.129\pm0.058$ &  88 & $ 0.0$ & 3\\  
$ -1.25$ & $  14.1\pm 10.6$ & $172.1\pm 7.2$ & $ 0.011\pm0.033$ & $-0.072\pm0.028$ &  88 & $ 0.0$ & 3\\  
$ -0.41$ & $   9.1\pm  8.4$ & $167.9\pm 5.5$ & $ 0.019\pm0.028$ & $-0.086\pm0.025$ &  88 & $ 0.0$ & 3\\  
$  0.20$ & $   2.7\pm  8.1$ & $178.1\pm 6.3$ & $ 0.018\pm0.030$ & $-0.058\pm0.023$ &  88 & $ 0.0$ & 3\\  
$  0.81$ & $   3.9\pm  8.4$ & $159.4\pm 9.2$ & $ 0.035\pm0.043$ & $-0.011\pm0.029$ &  88 & $ 0.0$ & 3\\  
$  1.65$ & $ -14.4\pm  8.1$ & $168.1\pm 9.6$ & $-0.002\pm0.027$ & $-0.070\pm0.031$ &  88 & $ 0.0$ & 3\\  
$  3.27$ & $ -26.9\pm 17.3$ & $133.8\pm20.2$ & $ 0.068\pm0.106$ & $-0.099\pm0.061$ &  88 & $ 0.0$ & 3\\ 
\cutinhead{GMP 3958} 
$ -2.94$ & $  -4.0\pm 26.7$ & $202.0\pm30.8$ & $ 0.069\pm0.079$ & $ 0.010\pm0.061$ &  12 & $ 0.0$ & 4\\ 
$ -1.82$ & $  12.6\pm 11.9$ & $157.4\pm13.5$ & $-0.007\pm0.045$ & $-0.063\pm0.033$ &  12 & $ 0.0$ & 4\\ 
$ -0.89$ & $  -4.1\pm  9.2$ & $161.0\pm 8.8$ & $ 0.048\pm0.035$ & $-0.117\pm0.032$ &  12 & $ 0.0$ & 4\\ 
$  0.03$ & $  -0.4\pm  6.7$ & $144.5\pm 6.9$ & $ 0.027\pm0.029$ & $-0.102\pm0.029$ &  12 & $ 0.0$ & 4\\ 
$  0.96$ & $  -7.5\pm  6.9$ & $143.3\pm 7.2$ & $ 0.038\pm0.030$ & $-0.086\pm0.029$ &  12 & $ 0.0$ & 4\\ 
$  1.88$ & $   5.6\pm 11.6$ & $160.8\pm10.5$ & $ 0.030\pm0.044$ & $-0.099\pm0.039$ &  12 & $ 0.0$ & 4\\ 
$  3.00$ & $  -1.9\pm 13.0$ & $145.5\pm15.3$ & $ 0.080\pm0.054$ & $-0.086\pm0.039$ &  12 & $ 0.0$ & 4\\ 
\cutinhead{GMP 4822}										         
$ -5.69$ & $  12.7\pm 15.8$ & $237.2\pm25.1$ & $ 0.089\pm0.056$ & $-0.005\pm0.053$ &  15 & $ 0.0$ & 5\\ 
$ -2.77$ & $  -4.2\pm  7.2$ & $240.2\pm 9.5$ & $ 0.029\pm0.023$ & $ 0.001\pm0.024$ &  15 & $ 0.0$ & 5\\ 
$ -1.34$ & $  -2.3\pm  4.9$ & $250.0\pm 7.1$ & $ 0.038\pm0.016$ & $ 0.019\pm0.017$ &  15 & $ 0.0$ & 5\\ 
$ -0.47$ & $   4.0\pm  4.9$ & $246.1\pm 6.4$ & $ 0.027\pm0.016$ & $-0.016\pm0.017$ &  15 & $ 0.0$ & 5\\ 
$  0.14$ & $   1.9\pm  4.5$ & $253.9\pm 6.1$ & $ 0.040\pm0.014$ & $-0.013\pm0.015$ &  15 & $ 0.0$ & 5\\ 
$  0.75$ & $   2.6\pm  5.1$ & $253.9\pm 7.5$ & $ 0.025\pm0.016$ & $-0.003\pm0.017$ &  15 & $ 0.0$ & 5\\ 
$  1.61$ & $  -3.2\pm  5.7$ & $254.4\pm 6.5$ & $ 0.012\pm0.016$ & $-0.009\pm0.018$ &  15 & $ 0.0$ & 5\\ 
$  3.06$ & $  -7.5\pm  6.6$ & $248.2\pm10.0$ & $ 0.029\pm0.021$ & $ 0.000\pm0.022$ &  15 & $ 0.0$ & 5\\ 
$  6.00$ & $  -3.7\pm 16.5$ & $253.9\pm18.6$ & $ 0.005\pm0.031$ & $ 0.009\pm0.036$ &  15 & $ 0.0$ & 5\\ 
\cutinhead{GMP 4928}										         
$ -3.64$ & $ -16.2\pm 45.3$ & $296.0\pm61.9$ & $ 0.046\pm0.100$ & $-0.024\pm0.078$ & 108 & $ 0.0$ & 2\\ 
$ -1.38$ & $  -0.8\pm 39.3$ & $334.5\pm63.7$ & $ 0.047\pm0.086$ & $ 0.111\pm0.102$ & 108 & $ 0.0$ & 2\\ 
$  0.05$ & $  19.5\pm 21.2$ & $307.1\pm36.8$ & $ 0.005\pm0.052$ & $ 0.060\pm0.048$ & 108 & $ 0.0$ & 2\\ 
$  1.48$ & $  18.8\pm 24.8$ & $293.5\pm26.1$ & $ 0.025\pm0.055$ & $-0.006\pm0.047$ & 108 & $ 0.0$ & 2\\ 
$  3.75$ & $ -21.2\pm 50.0$ & $317.5\pm69.3$ & $-0.079\pm0.099$ & $-0.003\pm0.068$ & 108 & $ 0.0$ & 2\\ 
$ -4.56$ & $  15.5\pm 32.7$ & $278.0\pm44.2$ & $-0.001\pm0.084$ & $-0.049\pm0.088$ & 108 & $ 0.0$ & 3\\  
$ -2.03$ & $  14.4\pm 14.3$ & $275.0\pm14.9$ & $ 0.023\pm0.038$ & $-0.021\pm0.034$ & 108 & $ 0.0$ & 3\\  
$ -0.60$ & $  -6.0\pm 11.8$ & $293.0\pm14.3$ & $ 0.005\pm0.030$ & $-0.002\pm0.032$ & 108 & $ 0.0$ & 3\\  
$  0.57$ & $  -1.8\pm  9.3$ & $293.5\pm11.2$ & $ 0.003\pm0.025$ & $-0.008\pm0.027$ & 108 & $ 0.0$ & 3\\  
$  2.00$ & $ -10.5\pm 14.0$ & $285.1\pm14.0$ & $ 0.037\pm0.041$ & $-0.042\pm0.041$ & 108 & $ 0.0$ & 3\\  
$  4.52$ & $ -11.4\pm 29.3$ & $277.5\pm36.4$ & $ 0.105\pm0.096$ & $-0.043\pm0.083$ & 108 & $ 0.0$ & 3\\  
$ -3.40$ & $ -36.7\pm 35.3$ & $292.4\pm48.5$ & $ 0.035\pm0.084$ & $-0.032\pm0.059$ & 153 & $ 0.0$ & 2\\ 
$ -2.45$ & $ -18.4\pm 15.2$ & $279.5\pm21.7$ & $-0.028\pm0.043$ & $-0.019\pm0.033$ & 153 & $ 0.0$ & 2\\ 
$ -1.58$ & $  14.2\pm 15.3$ & $282.9\pm23.2$ & $ 0.004\pm0.043$ & $-0.004\pm0.039$ & 153 & $ 0.0$ & 2\\ 
$ -0.97$ & $   1.5\pm 11.1$ & $301.5\pm17.5$ & $-0.003\pm0.033$ & $-0.033\pm0.032$ & 153 & $ 0.0$ & 2\\ 
$ -0.37$ & $  -6.7\pm 10.8$ & $288.6\pm16.7$ & $-0.018\pm0.033$ & $-0.020\pm0.028$ & 153 & $ 0.0$ & 2\\ 
$  0.24$ & $  12.0\pm 11.3$ & $295.2\pm17.7$ & $ 0.004\pm0.032$ & $-0.044\pm0.031$ & 153 & $ 0.0$ & 2\\ 
$  0.85$ & $  13.2\pm 12.6$ & $257.7\pm15.4$ & $-0.007\pm0.040$ & $-0.060\pm0.037$ & 153 & $ 0.0$ & 2\\ 
$  1.71$ & $  16.5\pm 15.0$ & $270.9\pm20.0$ & $ 0.034\pm0.040$ & $-0.023\pm0.036$ & 153 & $ 0.0$ & 2\\ 
$  2.93$ & $   4.7\pm 21.3$ & $267.0\pm28.1$ & $ 0.027\pm0.064$ & $-0.081\pm0.050$ & 153 & $ 0.0$ & 2\\ 
\cutinhead{GMP 5279} 
$ -9.26$ & $  -3.2\pm 32.3$ & $257.1\pm37.2$ & $-0.094\pm0.075$ & $-0.032\pm0.053$ & 146 & $ 0.0$ & 1\\ 
$ -2.78$ & $  -3.0\pm 11.2$ & $281.9\pm 7.7$ & $-0.009\pm0.031$ & $-0.066\pm0.028$ & 146 & $ 0.0$ & 1\\ 
$ -0.69$ & $  -1.4\pm  9.1$ & $280.8\pm 7.3$ & $-0.013\pm0.024$ & $-0.025\pm0.020$ & 146 & $ 0.0$ & 1\\ 
$  0.64$ & $  -1.1\pm 10.2$ & $292.2\pm 8.0$ & $-0.054\pm0.026$ & $-0.038\pm0.022$ & 146 & $ 0.0$ & 1\\ 
$  2.64$ & $   5.3\pm 10.3$ & $271.8\pm 7.9$ & $-0.009\pm0.024$ & $-0.049\pm0.026$ & 146 & $ 0.0$ & 1\\ 
$  8.38$ & $   3.7\pm 25.4$ & $246.7\pm26.5$ & $-0.027\pm0.075$ & $-0.092\pm0.072$ & 146 & $ 0.0$ & 1\\ 
\cutinhead{GMP 5568} 
$ -4.60$ & $   1.2\pm 25.0$ & $255.1\pm28.2$ & $-0.038\pm0.075$ & $-0.085\pm0.069$ &  78 & $ 2.8$ & 2\\ 
$ -2.05$ & $   2.1\pm 19.6$ & $264.6\pm19.6$ & $ 0.066\pm0.053$ & $ 0.003\pm0.047$ &  78 & $ 2.8$ & 2\\ 
$ -0.60$ & $  -1.1\pm 15.1$ & $263.9\pm19.3$ & $-0.005\pm0.040$ & $ 0.026\pm0.039$ &  78 & $ 2.8$ & 2\\ 
$  0.58$ & $ -18.2\pm 10.2$ & $228.2\pm15.2$ & $-0.005\pm0.043$ & $-0.014\pm0.038$ &  78 & $ 2.8$ & 2\\ 
$  2.03$ & $   6.0\pm 14.0$ & $235.4\pm19.3$ & $ 0.016\pm0.056$ & $-0.041\pm0.049$ &  78 & $ 2.8$ & 2\\ 
$  4.57$ & $  10.0\pm 33.4$ & $259.7\pm39.6$ & $ 0.026\pm0.079$ & $-0.080\pm0.067$ &  78 & $ 2.8$ & 2\\ 
$ -4.88$ & $ -12.1\pm133.3$ & $277.3\pm95.5$ & $-0.018\pm0.095$ & $-0.137\pm0.099$ &  78 & $14.0$ & 5\\ 
$ -1.26$ & $   8.6\pm 67.8$ & $259.0\pm98.0$ & $ 0.094\pm0.115$ & $-0.073\pm0.098$ &  78 & $14.0$ & 5\\ 
$  6.29$ & $  -8.6\pm109.4$ & $280.4\pm93.2$ & $ 0.077\pm0.153$ & $-0.081\pm0.136$ &  78 & $14.0$ & 5\\ 
$  2.66$ & $  12.3\pm 83.1$ & $317.1\pm67.1$ & $-0.029\pm0.120$ & $-0.064\pm0.102$ &  78 & $14.0$ & 5\\ 
$ -4.08$ & $   6.0\pm 29.8$ & $221.7\pm46.7$ & $-0.070\pm0.096$ & $ 0.004\pm0.068$ & 168 & $ 0.0$ & 2\\ 
$ -2.20$ & $  11.7\pm 13.8$ & $256.4\pm21.7$ & $ 0.023\pm0.047$ & $ 0.003\pm0.038$ & 168 & $ 0.0$ & 2\\ 
$ -1.36$ & $   2.3\pm 14.7$ & $257.2\pm21.0$ & $ 0.035\pm0.052$ & $-0.028\pm0.042$ & 168 & $ 0.0$ & 2\\ 
$ -0.75$ & $   9.4\pm  8.2$ & $248.7\pm10.8$ & $-0.014\pm0.034$ & $-0.040\pm0.029$ & 168 & $ 0.0$ & 2\\ 
$ -0.14$ & $   5.9\pm  7.0$ & $252.8\pm 9.5$ & $ 0.005\pm0.030$ & $-0.053\pm0.028$ & 168 & $ 0.0$ & 2\\ 
$  0.46$ & $   4.0\pm  9.5$ & $265.3\pm14.1$ & $-0.004\pm0.034$ & $-0.026\pm0.032$ & 168 & $ 0.0$ & 2\\ 
$  1.07$ & $ -14.2\pm 11.4$ & $240.1\pm16.3$ & $ 0.012\pm0.045$ & $-0.026\pm0.037$ & 168 & $ 0.0$ & 2\\ 
$  1.91$ & $  -3.3\pm 14.6$ & $237.6\pm21.8$ & $-0.025\pm0.055$ & $ 0.001\pm0.047$ & 168 & $ 0.0$ & 2\\ 
$  3.80$ & $ -21.5\pm 29.9$ & $234.1\pm40.5$ & $-0.060\pm0.103$ & $-0.081\pm0.066$ & 168 & $ 0.0$ & 2\\ 
$ -3.77$ & $   9.9\pm 25.5$ & $262.1\pm35.1$ & $ 0.031\pm0.054$ & $-0.039\pm0.070$ & 168 & $ 0.0$ & 5\\ 
$ -2.31$ & $   3.3\pm 12.6$ & $238.2\pm22.5$ & $-0.001\pm0.039$ & $ 0.023\pm0.047$ & 168 & $ 0.0$ & 5\\ 
$ -1.47$ & $  16.3\pm 13.2$ & $265.9\pm19.1$ & $-0.020\pm0.033$ & $-0.018\pm0.042$ & 168 & $ 0.0$ & 5\\ 
$ -0.86$ & $  10.2\pm 11.5$ & $253.0\pm16.4$ & $ 0.008\pm0.033$ & $-0.031\pm0.039$ & 168 & $ 0.0$ & 5\\ 
$ -0.26$ & $  -4.5\pm 10.7$ & $253.5\pm12.7$ & $-0.004\pm0.028$ & $-0.035\pm0.030$ & 168 & $ 0.0$ & 5\\ 
$  0.35$ & $  -0.8\pm 12.4$ & $260.5\pm13.6$ & $-0.003\pm0.027$ & $-0.026\pm0.033$ & 168 & $ 0.0$ & 5\\ 
$  0.96$ & $ -17.7\pm 12.0$ & $235.6\pm19.3$ & $-0.011\pm0.037$ & $ 0.004\pm0.041$ & 168 & $ 0.0$ & 5\\ 
$  1.80$ & $ -22.5\pm 19.9$ & $266.8\pm26.9$ & $-0.005\pm0.047$ & $-0.015\pm0.054$ & 168 & $ 0.0$ & 5\\ 
$  3.50$ & $   5.7\pm 28.8$ & $238.7\pm40.4$ & $ 0.011\pm0.057$ & $-0.002\pm0.078$ & 168 & $ 0.0$ & 5\\ 
\cutinhead{GMP 5975}											    
$ -8.13$ & $-104.4\pm 15.0$ & $155.7\pm12.4$ & $ 0.125\pm0.078$ & $-0.218\pm0.134$ &  68 & $ 0.0$ & 5\\ 
$ -3.89$ & $ -64.5\pm  9.1$ & $163.3\pm 9.7$ & $ 0.065\pm0.028$ & $-0.039\pm0.048$ &  68 & $ 0.0$ & 5\\ 
$ -2.01$ & $ -50.0\pm  5.8$ & $199.8\pm 6.5$ & $ 0.010\pm0.021$ & $-0.089\pm0.027$ &  68 & $ 0.0$ & 5\\ 
$ -1.17$ & $ -25.7\pm  4.1$ & $205.4\pm 5.7$ & $-0.032\pm0.014$ & $-0.056\pm0.019$ &  68 & $ 0.0$ & 5\\ 
$ -0.57$ & $ -13.4\pm  3.3$ & $207.1\pm 4.7$ & $-0.005\pm0.012$ & $-0.051\pm0.016$ &  68 & $ 0.0$ & 5\\ 
$  0.04$ & $   4.0\pm  2.9$ & $208.0\pm 4.1$ & $-0.028\pm0.010$ & $-0.043\pm0.014$ &  68 & $ 0.0$ & 5\\ 
$  0.65$ & $  33.1\pm  4.5$ & $201.7\pm 4.3$ & $-0.038\pm0.013$ & $-0.027\pm0.016$ &  68 & $ 0.0$ & 5\\ 
$  1.25$ & $  42.1\pm  4.0$ & $197.2\pm 5.4$ & $-0.007\pm0.015$ & $-0.048\pm0.020$ &  68 & $ 0.0$ & 5\\ 
$  2.09$ & $  45.4\pm 11.2$ & $213.1\pm16.1$ & $-0.028\pm0.035$ & $-0.054\pm0.032$ &  68 & $ 0.0$ & 5\\ 
$  3.99$ & $  65.4\pm  8.6$ & $221.6\pm11.8$ & $-0.016\pm0.027$ & $-0.063\pm0.025$ &  68 & $ 0.0$ & 5\\ 
$  7.82$ & $  67.9\pm 16.0$ & $190.4\pm20.8$ & $-0.016\pm0.061$ & $-0.151\pm0.062$ &  68 & $ 0.0$ & 5\\ 
\enddata 
\end{deluxetable} 
 
\clearpage 
 
\begin{deluxetable}{rcccccrrr} 
\tablecolumns{9} 
\tabletypesize{\scriptsize} 
\tablewidth{0pt} 
\tablecaption{Line strength indices of the sample galaxies \label{tab:indices}} 
\tablehead{ 
\colhead{$r$} & \colhead{\Hb} & \colhead{[MgFe]} &  
\colhead{\Fe} & \colhead{\Mgb} & \colhead{\Mgd} &  
\colhead{PA} & \colhead{Offset} & \colhead{Run}\\ 
\colhead{[arcsec]} & \colhead{[\AA]} & \colhead{[\AA]} &  
\colhead{[\AA]} & \colhead{[\AA]} & \colhead{[mag]} & 
\colhead{[$^\circ$]} & \colhead{[arcsec]} & \colhead{}} 
\startdata 
\cutinhead{GMP 0144} 
$ -3.83$ & $1.471\pm0.276$ & $3.414\pm0.284$ & $2.140\pm0.512$ & $5.446\pm0.325$ & $0.320\pm0.008$ &  46 & $ 0.0$ & 1\\ 
$ -0.88$ & $1.794\pm0.165$ & $3.696\pm0.110$ & $2.656\pm0.305$ & $5.141\pm0.196$ & $0.314\pm0.005$ &  46 & $ 0.0$ & 1\\ 
$  0.83$ & $1.330\pm0.175$ & $3.555\pm0.113$ & $2.560\pm0.312$ & $4.936\pm0.204$ & $0.308\pm0.005$ &  46 & $ 0.0$ & 1\\ 
$  3.90$ & $1.014\pm0.427$ & $3.589\pm0.663$ & $2.725\pm0.740$ & $4.727\pm0.499$ & $0.288\pm0.012$ &  46 & $ 0.0$ & 1\\ 
$-13.86$ & $2.131\pm0.305$ & $3.062\pm0.268$ & $2.802\pm0.494$ & $3.346\pm0.355$ & $0.213\pm0.009$ &   1 & $ 0.0$ & 1\\ 
$ -5.52$ & $1.989\pm0.191$ & $3.325\pm0.128$ & $2.507\pm0.337$ & $4.410\pm0.228$ & $0.262\pm0.006$ &   1 & $ 0.0$ & 1\\ 
$ -2.37$ & $1.652\pm0.132$ & $3.720\pm0.071$ & $2.982\pm0.242$ & $4.642\pm0.158$ & $0.288\pm0.004$ &   1 & $ 0.0$ & 1\\ 
$ -0.79$ & $1.729\pm0.120$ & $3.910\pm0.067$ & $2.923\pm0.234$ & $5.232\pm0.147$ & $0.292\pm0.004$ &   1 & $ 0.0$ & 1\\ 
$  0.57$ & $1.719\pm0.102$ & $3.846\pm0.044$ & $3.236\pm0.187$ & $4.571\pm0.123$ & $0.278\pm0.003$ &   1 & $ 0.0$ & 1\\ 
$  2.14$ & $1.444\pm0.113$ & $3.817\pm0.052$ & $3.138\pm0.203$ & $4.642\pm0.135$ & $0.274\pm0.003$ &   1 & $ 0.0$ & 1\\ 
$  5.27$ & $1.243\pm0.152$ & $3.695\pm0.092$ & $3.312\pm0.273$ & $4.123\pm0.183$ & $0.247\pm0.005$ &   1 & $ 0.0$ & 1\\ 
$ 13.71$ & $0.610\pm0.304$ & $3.451\pm0.323$ & $2.981\pm0.525$ & $3.995\pm0.357$ & $0.222\pm0.010$ &   1 & $ 0.0$ & 1\\	  
\cutinhead{GMP 2417} 
$ -6.92$ & $1.025\pm0.961$ & $3.189\pm1.436$ & $2.545\pm1.507$ & $3.996\pm1.231$ & $0.153\pm0.036$ &  55 & $-3.8$ & 3\\  
$ -2.56$ & $1.999\pm0.597$ & $3.417\pm1.007$ & $2.478\pm0.991$ & $4.713\pm0.893$ & $0.175\pm0.019$ &  55 & $-3.8$ & 3\\  
$  1.43$ & $1.292\pm0.553$ & $3.153\pm0.890$ & $3.007\pm0.914$ & $3.306\pm0.861$ & $0.169\pm0.021$ &  55 & $-3.8$ & 3\\  
$  6.23$ & $1.798\pm1.007$ & $3.278\pm1.320$ & $2.912\pm1.411$ & $3.690\pm1.185$ & $0.141\pm0.027$ &  55 & $-3.8$ & 3\\  
$ -3.29$ & $1.459\pm0.511$ & $3.279\pm0.799$ & $2.760\pm0.800$ & $3.896\pm0.768$ & $0.194\pm0.019$ & 145 & $ 0.0$ & 3\\  
$ -1.84$ & $2.145\pm0.234$ & $3.453\pm0.404$ & $2.685\pm0.408$ & $4.439\pm0.364$ & $0.233\pm0.007$ & 145 & $ 0.0$ & 3\\  
$ -1.00$ & $1.590\pm0.212$ & $3.495\pm0.354$ & $2.781\pm0.356$ & $4.391\pm0.329$ & $0.252\pm0.008$ & 145 & $ 0.0$ & 3\\  
$ -0.40$ & $1.408\pm0.155$ & $3.684\pm0.262$ & $2.961\pm0.267$ & $4.582\pm0.239$ & $0.259\pm0.005$ & 145 & $ 0.0$ & 3\\  
$  0.21$ & $1.455\pm0.163$ & $3.631\pm0.277$ & $2.901\pm0.282$ & $4.545\pm0.250$ & $0.263\pm0.006$ & 145 & $ 0.0$ & 3\\  
$  0.81$ & $1.597\pm0.185$ & $3.478\pm0.316$ & $2.720\pm0.321$ & $4.448\pm0.284$ & $0.259\pm0.006$ & 145 & $ 0.0$ & 3\\  
$  1.66$ & $1.604\pm0.250$ & $3.074\pm0.402$ & $2.312\pm0.421$ & $4.086\pm0.327$ & $0.240\pm0.008$ & 145 & $ 0.0$ & 3\\  
$  3.66$ & $0.344\pm0.463$ & $3.485\pm0.731$ & $2.549\pm0.715$ & $4.764\pm0.662$ & $0.229\pm0.014$ & 145 & $ 0.0$ & 3\\  
\cutinhead{GMP 2440} 
$ -3.00$ & $1.504\pm0.197$ & $3.093\pm0.108$ & $2.258\pm0.315$ & $4.236\pm0.222$ & $0.279\pm0.006$ &  75 & $ 0.0$ & 1\\  
$ -0.76$ & $1.764\pm0.143$ & $4.069\pm0.088$ & $3.044\pm0.256$ & $5.438\pm0.169$ & $0.299\pm0.005$ &  75 & $ 0.0$ & 1\\  
$  0.35$ & $1.974\pm0.132$ & $3.735\pm0.073$ & $2.813\pm0.245$ & $4.958\pm0.160$ & $0.297\pm0.004$ &  75 & $ 0.0$ & 1\\  
$  1.45$ & $1.431\pm0.217$ & $3.662\pm0.166$ & $2.881\pm0.365$ & $4.655\pm0.249$ & $0.290\pm0.007$ &  75 & $ 0.0$ & 1\\  
$  4.27$ & $0.832\pm0.307$ & $3.946\pm0.329$ & $3.597\pm0.479$ & $4.330\pm0.348$ & $0.284\pm0.009$ &  75 & $ 0.0$ & 1\\  
$ -4.36$ & $1.331\pm0.301$ & $3.030\pm0.256$ & $2.343\pm0.495$ & $3.918\pm0.342$ & $0.261\pm0.009$ &  18 & $ 0.0$ & 1\\  
$ -1.15$ & $1.524\pm0.156$ & $3.787\pm0.095$ & $2.962\pm0.274$ & $4.842\pm0.183$ & $0.296\pm0.005$ &  18 & $ 0.0$ & 1\\  
$  0.88$ & $1.463\pm0.130$ & $3.625\pm0.064$ & $2.950\pm0.230$ & $4.456\pm0.154$ & $0.285\pm0.004$ &  18 & $ 0.0$ & 1\\  
$  4.06$ & $1.323\pm0.346$ & $3.287\pm0.359$ & $2.829\pm0.550$ & $3.819\pm0.397$ & $0.269\pm0.011$ &  18 & $ 0.0$ & 1\\  
\cutinhead{GMP 2921}										         
$-11.08$ & $2.430\pm0.411$ & $3.732\pm0.913$ & $2.872\pm1.009$ & $4.848\pm0.670$ & $0.293\pm0.016$ & 171 & $ 0.0$ & 5\\  
$ -5.90$ & $1.692\pm0.137$ & $3.841\pm0.268$ & $2.951\pm0.298$ & $4.999\pm0.193$ & $0.316\pm0.004$ & 171 & $ 0.0$ & 5\\  
$ -2.60$ & $1.590\pm0.104$ & $4.156\pm0.222$ & $3.197\pm0.244$ & $5.403\pm0.165$ & $0.347\pm0.004$ & 171 & $ 0.0$ & 5\\  
$ -1.16$ & $1.532\pm0.075$ & $4.158\pm0.163$ & $3.070\pm0.167$ & $5.631\pm0.134$ & $0.364\pm0.003$ & 171 & $ 0.0$ & 5\\  
$  0.03$ & $1.584\pm0.066$ & $4.063\pm0.143$ & $3.049\pm0.155$ & $5.413\pm0.105$ & $0.366\pm0.003$ & 171 & $ 0.0$ & 5\\  
$  1.21$ & $1.582\pm0.073$ & $4.079\pm0.159$ & $3.108\pm0.162$ & $5.353\pm0.139$ & $0.368\pm0.003$ & 171 & $ 0.0$ & 5\\  
$  2.65$ & $1.507\pm0.092$ & $4.032\pm0.197$ & $3.000\pm0.214$ & $5.419\pm0.143$ & $0.367\pm0.004$ & 171 & $ 0.0$ & 5\\  
$  5.76$ & $1.948\pm0.123$ & $4.169\pm0.277$ & $3.016\pm0.294$ & $5.761\pm0.203$ & $0.349\pm0.005$ & 171 & $ 0.0$ & 5\\  
$ 10.76$ & $2.100\pm0.338$ & $4.030\pm0.769$ & $2.736\pm0.798$ & $5.936\pm0.535$ & $0.318\pm0.013$ & 171 & $ 0.0$ & 5\\  
\cutinhead{GMP 3329}										         
$ -9.50$ & $1.847\pm0.475$ & $3.781\pm0.845$ & $2.853\pm0.908$ & $5.010\pm0.645$ & $0.299\pm0.018$ & 135 & $ 0.0$ & 5\\ 
$ -4.86$ & $1.807\pm0.231$ & $3.481\pm0.398$ & $2.644\pm0.427$ & $4.583\pm0.307$ & $0.300\pm0.009$ & 135 & $ 0.0$ & 5\\ 
$ -2.55$ & $1.303\pm0.145$ & $3.653\pm0.257$ & $2.770\pm0.277$ & $4.816\pm0.196$ & $0.312\pm0.005$ & 135 & $ 0.0$ & 5\\ 
$ -1.08$ & $1.613\pm0.139$ & $3.580\pm0.248$ & $2.704\pm0.267$ & $4.740\pm0.189$ & $0.309\pm0.005$ & 135 & $ 0.0$ & 5\\ 
$  0.11$ & $1.499\pm0.141$ & $3.574\pm0.253$ & $2.663\pm0.261$ & $4.796\pm0.209$ & $0.322\pm0.005$ & 135 & $ 0.0$ & 5\\ 
$  1.31$ & $1.330\pm0.145$ & $3.627\pm0.257$ & $2.746\pm0.278$ & $4.791\pm0.196$ & $0.321\pm0.005$ & 135 & $ 0.0$ & 5\\ 
$  2.78$ & $1.410\pm0.164$ & $3.586\pm0.271$ & $2.730\pm0.295$ & $4.711\pm0.202$ & $0.315\pm0.007$ & 135 & $ 0.0$ & 5\\ 
$  5.08$ & $1.004\pm0.257$ & $3.184\pm0.438$ & $2.265\pm0.456$ & $4.477\pm0.329$ & $0.310\pm0.010$ & 135 & $ 0.0$ & 5\\ 
$  9.56$ & $1.371\pm0.416$ & $3.309\pm0.754$ & $2.377\pm0.762$ & $4.607\pm0.622$ & $0.281\pm0.013$ & 135 & $ 0.0$ & 5\\ 
\cutinhead{GMP 3414}										         
$ -6.88$ & $3.114\pm0.797$ & $4.175\pm1.204$ & $3.861\pm1.256$ & $4.515\pm1.135$ & $0.218\pm0.036$ & 178 & $ 4.6$ & 3\\ 
$ -2.47$ & $2.941\pm0.467$ & $4.273\pm0.715$ & $3.946\pm0.734$ & $4.628\pm0.687$ & $0.205\pm0.015$ & 178 & $ 4.6$ & 3\\ 
$  1.56$ & $3.037\pm0.443$ & $4.040\pm0.742$ & $3.466\pm0.728$ & $4.710\pm0.741$ & $0.275\pm0.019$ & 178 & $ 4.6$ & 3\\ 
$  6.01$ & $3.999\pm0.642$ & $3.195\pm0.992$ & $2.692\pm1.002$ & $3.791\pm0.943$ & $0.290\pm0.021$ & 178 & $ 4.6$ & 3\\ 
$ -2.99$ & $0.944\pm0.318$ & $3.357\pm0.463$ & $2.694\pm0.492$ & $4.183\pm0.389$ & $0.206\pm0.011$ &  88 & $ 0.0$ & 3\\ 
$ -1.25$ & $1.872\pm0.125$ & $3.448\pm0.196$ & $2.751\pm0.198$ & $4.321\pm0.182$ & $0.233\pm0.004$ &  88 & $ 0.0$ & 3\\ 
$ -0.41$ & $1.700\pm0.103$ & $3.425\pm0.160$ & $2.798\pm0.162$ & $4.193\pm0.149$ & $0.260\pm0.003$ &  88 & $ 0.0$ & 3\\ 
$  0.20$ & $1.342\pm0.097$ & $3.444\pm0.151$ & $2.853\pm0.153$ & $4.156\pm0.141$ & $0.269\pm0.003$ &  88 & $ 0.0$ & 3\\ 
$  0.81$ & $1.261\pm0.121$ & $3.346\pm0.185$ & $2.689\pm0.186$ & $4.164\pm0.173$ & $0.259\pm0.004$ &  88 & $ 0.0$ & 3\\ 
$  1.65$ & $1.680\pm0.132$ & $3.371\pm0.211$ & $2.508\pm0.208$ & $4.531\pm0.192$ & $0.254\pm0.004$ &  88 & $ 0.0$ & 3\\ 
$  3.27$ & $2.480\pm0.290$ & $2.956\pm0.484$ & $1.928\pm0.454$ & $4.532\pm0.418$ & $0.246\pm0.009$ &  88 & $ 0.0$ & 3\\ 
\cutinhead{GMP 3958} 
$ -2.94$ & $2.359\pm0.408$ & $2.905\pm0.738$ & $2.720\pm0.741$ & $3.102\pm0.686$ & $0.221\pm0.015$ &  12 & $ 0.0$ & 4\\    
$ -1.82$ & $1.712\pm0.302$ & \nodata         & $2.516\pm0.508$ & \nodata         & \nodata         &  12 & $ 0.0$ & 4\\    
$ -0.89$ & $1.519\pm0.220$ & \nodata         & $2.350\pm0.372$ & \nodata         & \nodata         &  12 & $ 0.0$ & 4\\     
$  0.03$ & $2.135\pm0.196$ & \nodata         & $2.298\pm0.322$ & \nodata         & \nodata         &  12 & $ 0.0$ & 4\\    
$  0.96$ & $1.605\pm0.198$ & $2.873\pm0.150$ & $2.351\pm0.331$ & $3.510\pm0.314$ & $0.274\pm0.007$ &  12 & $ 0.0$ & 4\\    
$  1.88$ & $2.270\pm0.319$ & $3.164\pm0.406$ & $2.479\pm0.560$ & $4.039\pm0.458$ & $0.268\pm0.012$ &  12 & $ 0.0$ & 4\\    
$  3.00$ & $0.768\pm0.399$ & $3.332\pm0.671$ & $2.695\pm0.650$ & $4.120\pm0.620$ & $0.260\pm0.015$ &  12 & $ 0.0$ & 4\\    
\cutinhead{GMP 4822}										         
$ -5.69$ & $1.462\pm0.295$ & $3.402\pm0.489$ & $2.760\pm0.540$ & $4.193\pm0.386$ & $0.271\pm0.011$ &  15 & $ 0.0$ & 5\\ 
$ -2.77$ & $1.226\pm0.138$ & $3.542\pm0.235$ & $2.691\pm0.237$ & $4.663\pm0.208$ & $0.290\pm0.005$ &  15 & $ 0.0$ & 5\\ 
$ -1.34$ & $1.561\pm0.090$ & $3.608\pm0.156$ & $2.726\pm0.167$ & $4.776\pm0.120$ & $0.311\pm0.003$ &  15 & $ 0.0$ & 5\\ 
$ -0.47$ & $1.751\pm0.095$ & $3.498\pm0.155$ & $2.560\pm0.164$ & $4.779\pm0.116$ & $0.335\pm0.003$ &  15 & $ 0.0$ & 5\\ 
$  0.14$ & $1.490\pm0.084$ & $3.627\pm0.150$ & $2.630\pm0.158$ & $5.001\pm0.113$ & $0.336\pm0.003$ &  15 & $ 0.0$ & 5\\ 
$  0.75$ & $1.501\pm0.097$ & $3.715\pm0.160$ & $2.839\pm0.171$ & $4.862\pm0.126$ & $0.334\pm0.003$ &  15 & $ 0.0$ & 5\\ 
$  1.61$ & $1.530\pm0.101$ & $3.638\pm0.173$ & $2.808\pm0.187$ & $4.713\pm0.135$ & $0.329\pm0.004$ &  15 & $ 0.0$ & 5\\ 
$  3.06$ & $1.244\pm0.132$ & $3.462\pm0.226$ & $2.596\pm0.230$ & $4.618\pm0.193$ & $0.313\pm0.005$ &  15 & $ 0.0$ & 5\\ 
$  6.00$ & $1.931\pm0.240$ & $3.466\pm0.421$ & $2.574\pm0.448$ & $4.667\pm0.321$ & $0.296\pm0.009$ &  15 & $ 0.0$ & 5\\ 
\cutinhead{GMP 4928}										         
$ -3.64$ & $1.248\pm0.331$ & $4.128\pm1.160$ & $3.215\pm0.844$ & $5.301\pm0.666$ & $0.277\pm0.017$ & 108 & $ 0.0$ & 2\\     
$ -1.38$ & $1.096\pm0.203$ & $4.375\pm0.424$ & $3.807\pm0.568$ & $5.029\pm0.341$ & $0.316\pm0.010$ & 108 & $ 0.0$ & 2\\     
$  0.05$ & $1.370\pm0.142$ & $4.382\pm0.238$ & $3.498\pm0.372$ & $5.488\pm0.292$ & $0.320\pm0.007$ & 108 & $ 0.0$ & 2\\ 
$  1.48$ & $0.908\pm0.161$ & $3.775\pm0.187$ & $2.859\pm0.399$ & $4.985\pm0.249$ & $0.317\pm0.008$ & 108 & $ 0.0$ & 2\\ 
$  3.75$ & $2.386\pm0.292$ & $3.478\pm0.845$ & $3.398\pm0.791$ & $3.559\pm0.615$ & $0.296\pm0.015$ & 108 & $ 0.0$ & 2\\ 
$ -4.56$ & $1.667\pm0.427$ & $3.742\pm0.775$ & $3.440\pm0.789$ & $4.070\pm0.752$ & $0.230\pm0.019$ & 108 & $ 0.0$ & 3\\ 
$ -2.03$ & $1.641\pm0.198$ & $3.196\pm0.367$ & $2.377\pm0.369$ & $4.298\pm0.319$ & $0.259\pm0.006$ & 108 & $ 0.0$ & 3\\ 
$ -0.60$ & $1.199\pm0.171$ & $3.708\pm0.317$ & $2.888\pm0.321$ & $4.760\pm0.285$ & $0.277\pm0.007$ & 108 & $ 0.0$ & 3\\ 
$  0.57$ & $1.268\pm0.139$ & $3.772\pm0.255$ & $2.988\pm0.262$ & $4.762\pm0.226$ & $0.286\pm0.004$ & 108 & $ 0.0$ & 3\\ 
$  2.00$ & $1.037\pm0.212$ & $3.746\pm0.376$ & $3.230\pm0.389$ & $4.344\pm0.349$ & $0.283\pm0.008$ & 108 & $ 0.0$ & 3\\ 
$  4.52$ & $1.137\pm0.366$ & $3.433\pm0.665$ & $2.597\pm0.671$ & $4.538\pm0.584$ & $0.256\pm0.011$ & 108 & $ 0.0$ & 3\\ 
$ -3.40$ & $0.403\pm0.248$ & $3.682\pm0.558$ & $2.974\pm0.619$ & $4.559\pm0.490$ & $0.270\pm0.013$ & 153 & $ 0.0$ & 2\\ 
$ -2.45$ & $1.476\pm0.126$ & $3.388\pm0.132$ & $2.540\pm0.312$ & $4.520\pm0.249$ & $0.282\pm0.007$ & 153 & $ 0.0$ & 2\\ 
$ -1.58$ & $2.145\pm0.129$ & $3.420\pm0.144$ & $2.488\pm0.326$ & $4.699\pm0.259$ & $0.286\pm0.007$ & 153 & $ 0.0$ & 2\\ 
$ -0.97$ & $2.032\pm0.091$ & $3.850\pm0.084$ & $3.015\pm0.236$ & $4.915\pm0.186$ & $0.290\pm0.005$ & 153 & $ 0.0$ & 2\\ 
$ -0.37$ & $1.544\pm0.091$ & $3.719\pm0.078$ & $2.929\pm0.230$ & $4.722\pm0.183$ & $0.289\pm0.005$ & 153 & $ 0.0$ & 2\\ 
$  0.24$ & $1.600\pm0.096$ & $3.675\pm0.082$ & $2.571\pm0.225$ & $5.253\pm0.199$ & $0.295\pm0.005$ & 153 & $ 0.0$ & 2\\ 
$  0.85$ & $1.438\pm0.121$ & $3.257\pm0.110$ & $2.418\pm0.289$ & $4.388\pm0.234$ & $0.299\pm0.006$ & 153 & $ 0.0$ & 2\\ 
$  1.71$ & $1.390\pm0.127$ & $3.577\pm0.131$ & $2.571\pm0.294$ & $4.978\pm0.249$ & $0.311\pm0.007$ & 153 & $ 0.0$ & 2\\ 
$  2.93$ & $0.902\pm0.191$ & $3.114\pm0.266$ & $2.366\pm0.460$ & $4.098\pm0.371$ & $0.292\pm0.010$ & 153 & $ 0.0$ & 2\\ 
\cutinhead{GMP 5279} 
$ -9.26$ & $1.396\pm0.455$ & $2.751\pm0.702$ & $2.559\pm0.884$ & $2.958\pm0.577$ & $0.190\pm0.013$ & 146 & $ 0.0$ & 1\\ 
$ -2.78$ & $1.711\pm0.152$ & $3.729\pm0.112$ & $3.023\pm0.309$ & $4.599\pm0.194$ & $0.282\pm0.004$ & 146 & $ 0.0$ & 1\\ 
$ -0.69$ & $1.805\pm0.121$ & $3.781\pm0.072$ & $3.035\pm0.246$ & $4.711\pm0.154$ & $0.304\pm0.004$ & 146 & $ 0.0$ & 1\\ 
$  0.64$ & $1.473\pm0.125$ & $4.379\pm0.090$ & $3.765\pm0.257$ & $5.092\pm0.160$ & $0.303\pm0.004$ & 146 & $ 0.0$ & 1\\ 
$  2.64$ & $1.163\pm0.146$ & $3.872\pm0.102$ & $3.119\pm0.289$ & $4.808\pm0.183$ & $0.281\pm0.004$ & 146 & $ 0.0$ & 1\\ 
$  8.38$ & $2.002\pm0.509$ & $3.274\pm1.016$ & $2.899\pm0.971$ & $3.696\pm0.639$ & $0.211\pm0.016$ & 146 & $ 0.0$ & 1\\ 
\cutinhead{GMP 5568} 
$ -4.60$ & $1.301\pm0.251$ & $3.357\pm0.485$ & $2.650\pm0.597$ & $4.252\pm0.484$ & $0.278\pm0.013$ &  78 & $ 2.8$ & 2\\ 
$ -2.05$ & $0.972\pm0.153$ & $3.676\pm0.201$ & $2.897\pm0.368$ & $4.665\pm0.297$ & $0.306\pm0.008$ &  78 & $ 2.8$ & 2\\ 
$ -0.60$ & $1.532\pm0.122$ & $4.007\pm0.143$ & $3.307\pm0.298$ & $4.855\pm0.240$ & $0.312\pm0.006$ &  78 & $ 2.8$ & 2\\ 
$  0.58$ & $1.483\pm0.115$ & $3.435\pm0.080$ & $2.465\pm0.229$ & $4.787\pm0.204$ & $0.283\pm0.005$ &  78 & $ 2.8$ & 2\\ 
$  2.03$ & $0.982\pm0.160$ & $3.390\pm0.186$ & $2.470\pm0.365$ & $4.651\pm0.300$ & $0.276\pm0.008$ &  78 & $ 2.8$ & 2\\ 
$  4.57$ & $1.210\pm0.254$ & $3.248\pm0.446$ & $2.769\pm0.559$ & $3.811\pm0.491$ & $0.239\pm0.012$ &  78 & $ 2.8$ & 2\\ 
$ -4.88$ & $3.543\pm0.527$ & $3.731\pm1.230$ & $2.546\pm1.265$ & $5.468\pm0.889$ & $0.287\pm0.026$ &  78 & $14.0$ & 5\\ 
$ -1.26$ & $1.359\pm0.674$ & $4.238\pm1.447$ & $3.016\pm1.515$ & $5.956\pm1.074$ & $0.299\pm0.032$ &  78 & $14.0$ & 5\\ 
$  2.66$ & $3.033\pm0.520$ & $3.310\pm1.386$ & $1.987\pm1.333$ & $5.516\pm0.915$ & $0.283\pm0.025$ &  78 & $14.0$ & 5\\ 
$  6.29$ & $1.979\pm0.871$ & $3.692\pm1.847$ & $3.079\pm2.072$ & $4.427\pm1.450$ & $0.286\pm0.042$ &  78 & $14.0$ & 5\\ 
$ -4.08$ & $0.434\pm0.271$ & $3.096\pm0.468$ & $2.619\pm0.605$ & $3.661\pm0.500$ & $0.248\pm0.014$ & 168 & $ 0.0$ & 2\\ 
$ -2.20$ & $1.313\pm0.135$ & $3.768\pm0.157$ & $2.752\pm0.321$ & $5.157\pm0.260$ & $0.265\pm0.007$ & 168 & $ 0.0$ & 2\\ 
$ -1.36$ & $1.300\pm0.145$ & $3.798\pm0.152$ & $2.826\pm0.303$ & $5.103\pm0.265$ & $0.268\pm0.006$ & 168 & $ 0.0$ & 2\\ 
$ -0.75$ & $1.373\pm0.089$ & $3.643\pm0.065$ & $2.680\pm0.210$ & $4.952\pm0.171$ & $0.280\pm0.005$ & 168 & $ 0.0$ & 2\\ 
$ -0.14$ & $1.343\pm0.079$ & $3.845\pm0.050$ & $3.013\pm0.172$ & $4.908\pm0.152$ & $0.287\pm0.004$ & 168 & $ 0.0$ & 2\\ 
$  0.46$ & $1.272\pm0.095$ & $3.868\pm0.083$ & $2.895\pm0.231$ & $5.170\pm0.186$ & $0.296\pm0.005$ & 168 & $ 0.0$ & 2\\ 
$  1.07$ & $1.405\pm0.126$ & $3.408\pm0.112$ & $2.477\pm0.306$ & $4.688\pm0.214$ & $0.290\pm0.006$ & 168 & $ 0.0$ & 2\\ 
$  1.91$ & $0.810\pm0.157$ & $3.340\pm0.176$ & $2.373\pm0.359$ & $4.701\pm0.294$ & $0.300\pm0.008$ & 168 & $ 0.0$ & 2\\ 
$  3.80$ & $1.333\pm0.302$ & $2.913\pm0.549$ & $2.053\pm0.677$ & $4.135\pm0.557$ & $0.265\pm0.016$ & 168 & $ 0.0$ & 2\\ 
$ -3.77$ & $1.503\pm0.412$ & $3.354\pm0.730$ & $2.504\pm0.781$ & $4.492\pm0.553$ & $0.257\pm0.016$ & 168 & $ 0.0$ & 5\\ 
$ -2.31$ & $1.677\pm0.253$ & $3.380\pm0.423$ & $2.495\pm0.447$ & $4.579\pm0.325$ & $0.265\pm0.009$ & 168 & $ 0.0$ & 5\\ 
$ -1.47$ & $1.583\pm0.229$ & $3.710\pm0.400$ & $2.729\pm0.424$ & $5.043\pm0.304$ & $0.274\pm0.009$ & 168 & $ 0.0$ & 5\\ 
$ -0.86$ & $1.564\pm0.228$ & $3.605\pm0.386$ & $2.713\pm0.423$ & $4.792\pm0.279$ & $0.280\pm0.008$ & 168 & $ 0.0$ & 5\\ 
$ -0.26$ & $1.631\pm0.193$ & $3.900\pm0.333$ & $2.901\pm0.356$ & $5.242\pm0.251$ & $0.295\pm0.007$ & 168 & $ 0.0$ & 5\\ 
$  0.35$ & $1.575\pm0.202$ & $3.671\pm0.378$ & $2.692\pm0.392$ & $5.005\pm0.302$ & $0.275\pm0.006$ & 168 & $ 0.0$ & 5\\ 
$  0.96$ & $1.295\pm0.242$ & $3.647\pm0.425$ & $2.512\pm0.433$ & $5.294\pm0.320$ & $0.274\pm0.009$ & 168 & $ 0.0$ & 5\\ 
$  1.80$ & $1.881\pm0.296$ & $3.624\pm0.566$ & $2.404\pm0.569$ & $5.465\pm0.415$ & $0.268\pm0.011$ & 168 & $ 0.0$ & 5\\ 
$  3.50$ & $1.687\pm0.552$ & $3.326\pm0.907$ & $2.467\pm0.973$ & $4.485\pm0.678$ & $0.264\pm0.020$ & 168 & $ 0.0$ & 5\\ 
\cutinhead{GMP 5975}											    
$ -8.13$ & $0.863\pm0.450$ & $2.885\pm0.628$ & $2.432\pm0.692$ & $3.421\pm0.515$ & $0.195\pm0.013$ &  68 & $ 0.0$ & 5\\	    
$ -3.89$ & $1.553\pm0.211$ & $3.078\pm0.310$ & $2.724\pm0.346$ & $3.479\pm0.259$ & $0.211\pm0.008$ &  68 & $ 0.0$ & 5\\	    
$ -2.01$ & $1.724\pm0.140$ & $3.392\pm0.210$ & $2.855\pm0.227$ & $4.029\pm0.179$ & $0.215\pm0.004$ &  68 & $ 0.0$ & 5\\	    
$ -1.17$ & $1.766\pm0.102$ & $3.488\pm0.161$ & $2.833\pm0.176$ & $4.294\pm0.130$ & $0.249\pm0.004$ &  68 & $ 0.0$ & 5\\	    
$ -0.57$ & $1.671\pm0.084$ & $3.508\pm0.136$ & $2.787\pm0.150$ & $4.414\pm0.105$ & $0.298\pm0.003$ &  68 & $ 0.0$ & 5\\	    
$  0.04$ & $1.609\pm0.071$ & $3.539\pm0.114$ & $2.796\pm0.124$ & $4.479\pm0.091$ & $0.316\pm0.003$ &  68 & $ 0.0$ & 5\\	    
$  0.65$ & $1.767\pm0.082$ & $3.534\pm0.131$ & $2.812\pm0.140$ & $4.440\pm0.109$ & $0.307\pm0.003$ &  68 & $ 0.0$ & 5\\	    
$  1.25$ & $1.749\pm0.105$ & $3.561\pm0.165$ & $2.920\pm0.180$ & $4.343\pm0.134$ & $0.290\pm0.004$ &  68 & $ 0.0$ & 5\\	    
$  2.09$ & $1.594\pm0.263$ & $3.512\pm0.388$ & $2.995\pm0.431$ & $4.119\pm0.318$ & $0.279\pm0.010$ &  68 & $ 0.0$ & 5\\ 
$  3.99$ & $1.824\pm0.193$ & $3.183\pm0.311$ & $2.667\pm0.346$ & $3.800\pm0.250$ & $0.246\pm0.007$ &  68 & $ 0.0$ & 5\\ 
$  7.82$ & $1.413\pm0.431$ & $2.760\pm0.645$ & $2.381\pm0.699$ & $3.200\pm0.555$ & $0.248\pm0.017$ &  68 & $ 0.0$ & 5\\ 
\enddata 
\end{deluxetable} 
 
\end{document}